\pdfoutput=1
\documentclass[12pt,aps,preprint,superscriptaddress,longbibliography]{revtex4-1}

\usepackage{amsmath}
\usepackage{amssymb}
\usepackage{mathrsfs}
\usepackage{graphicx}
\usepackage{dsfont}
\usepackage[pdftex,colorlinks=true,linkcolor=black,citecolor=black,urlcolor=black]{hyperref}
\usepackage{enumerate}
\usepackage{subcaption}
\usepackage{comment}

\usepackage{slashed}

\def\g{\gamma}

\def\d{\partial}

\def\cN{\mathcal{N}}

\def\cA{\mathcal{A}}
\def\cD{\mathcal{D}}
\def\cO{\mathcal{O}}

\def\cS{\mathcal{S}}
\def\cN{\mathcal{N}}

\def\sl2r{SL(2,\mathbb{R})}

\def\ce{\varepsilon}

\newcommand{\lsim}{\mathrel{\hbox{\rlap{\lower.55ex \hbox{$\sim$}} \kern-.3em \raise.4ex \hbox{$<$}}}}
\newcommand{\gsim}{\mathrel{\hbox{\rlap{\lower.55ex \hbox{$\sim$}} \kern-.3em \raise.4ex \hbox{$>$}}}}

\newcommand{\be}{\begin{equation}}
\newcommand{\ee}{\end{equation}}
\renewcommand\[{\begin{equation}}
\renewcommand\]{\end{equation}}

\newcommand{\sign}{\mathrm{sgn}} 

\usepackage{bm}

\def\eq#1 { \begin{equation} #1 \end{equation} }

\def\g{\gamma}

\def\n{\nu}
\def\w{\omega}

\def\w{\omega}

\begin{document}

\preprint{EFI-17-18}
\title{Particle-hole symmetry and composite fermions\\ in fractional quantum Hall states}
\author{Dung Xuan Nguyen}
\affiliation{Kadanoff Center for Theoretical Physics, University of Chicago, Chicago, Illinois 60637, USA}
\author{Siavash Golkar}
\affiliation{Center for Cosmology and Particle Physics, Department of Physics, New York University, New York, New York 10003, USA\\ $~$}
\author{Matthew M. Roberts}
\affiliation{Kadanoff Center for Theoretical Physics, University of Chicago, Chicago, Illinois 60637, USA}
\author{Dam Thanh Son}
\affiliation{Kadanoff Center for Theoretical Physics, University of Chicago, Chicago, Illinois 60637, USA}

\

\begin{abstract}
We study fractional quantum Hall states at filling fractions in the Jain sequences using the framework of composite Dirac fermions.    Synthesizing previous work, we write down an effective field theory consistent with all symmetry requirements, including Galilean invariance and particle-hole symmetry.  Employing a Fermi liquid description, we demonstrate the appearance of the Girvin--Macdonlald--Platzman algebra and compute the dispersion relation of neutral excitations and various response functions. Our results satisfy requirements of particle-hole symmetry.  We show that while the dispersion relation obtained from the HLR theory is particle-hole symmetric,  correlation functions obtained from HLR are not.  The results of the Dirac theory are shown to be consistent with the Haldane bound on the projected structure factor, while those of the HLR theory violate it. 
\end{abstract}

\maketitle
		
	\section{Introduction}

Since the discovery of the fractional quantum Hall effect (FQHE) \cite{Tsui:1982yy}, a vast amount of theoretical and experimental work has been done to explore this fascinating phenomenon. In spite of the effort, the FQHE remains one of the most challenging problems of condensed matter physics. A major breakthrough in approaching the problem was the idea of the composite fermion~\cite{Jain:1989tx,Jain:1992rs}, derived from an earlier flux-attachment approach \cite{Zhang:1988wy}. Starting from a mean field approximation, one arrives at an effective field theory of FQH, known as  Chern-Simons (CS) fermionic theory which was first used to describe Jain's sequence of incompressible fractionally quantized Hall states \cite{Fradkin:1991wy}.
	
The CS fermionic field theory was later used by Halperin, Lee, and Read (commonly referred to as HLR theory) \cite{Halperin:1992mh} to describe FQH states at filling fractions with the even denominators. A crucial test of the HLR theory was its prediction of compressible FQH states where the composite fermion forms a Fermi-liquid. This behavior of composite fermions near half-filling was consequently confirmed experimentally \cite{Willett:1990,Kang:1993imw,Goldman:1994zz} and constitutes one of the greatest triumphs of HLR theory, lending more credence to the idea that the physical degrees of freedom of FQH systems near half-filling are indeed composite fermions.
	
	
Despite its phenomenological success, the HLR theory and the flux attachment approach to FQH systems in general have been criticized on various grounds (see, e.g., Ref.~\cite{Dyakonov:2002}).  The most commonly raised criticisms---the wrong energy scale and the lack of projection to the lowest Landau level---can be partially addressed by phenomenological modifications of the HLR theory, the most successful of which is perhaps the ``magnetic modified random-phase approximation," or MMRPA \cite{MMRPA} (As its name reveals, the MMRPA in fact includes two separate modifications to HLR.  The first addresses the issue of the problem of wrong energy scale~\cite{Simon:1993}, and the second ensures the finiteness of physical observables in for massless limit of electrons with $g$-factor $g=2$, a property of the lowest Landau level~\cite{MMRPA}.)  The proposed modifications of HLR do not, however, address the issue of particle-hole symmetry (PHS) \cite{Girvin:ph}, which has again attracted attention after intriguing experimental results \cite{Kamburov:2014,Kamburov:2015} which indicate that the composite fermion density is less than the electron density if $\nu>1/2$, in contrast with the expectation from the HLR theory. 
	
The PHS is a focus of this paper.
In the lowest Landau level limit, with only two-body interaction, the projected Hamiltonian has this symmetry.  
Particle-hole conjugation maps a FQH state with filling factor $\nu$ to another state with filling factor $1-\nu$, and PHS imposes stringent constraints on physical observables in the two states. For example,
the projected density-density interaction is invariant under particle-hole conjugation. 
There are a more subtle relationship between the finite-wave-vector Hall conductivities of the two states \cite{SonLevin:PH}. Though the symmetry is realized only in the limit of very high magnetic field, any theory pretending to describe the quantum Hall effect should be capable of accommodating the symmetry. There is also some experimental evidence that particle-hole symmetry is relevant in real experiments \cite{Kamburov:2014,Kamburov:2015}. 
	
The fact that the process of flux attachment breaks particle-hole symmetry by attaching magnetic fluxes are to particles, but not holes, has received early attention. Kivelson, Lee, Krotov and Gan \cite{Kivelson:1997,Lee:1999} performed a simple calculation of the conductivity tensor $\sigma_{ij}$ using a random-phase approximation (RPA) of the HLR theory and obtained a result which does not satisfy constraints implied by particle-hole symmetry. The lack of explicit PHS of HLR theory is also apparent in the asymmetric treatment of Jain's sequences states, where states with filling fraction $\nu=\frac{N}{2N+1}$ and  $\nu=\frac{N+1}{2N+1}$ are mapped to different CF states with filling fractions $\nu_{CF}=N$ and  $\nu_{CF}=N+1$ respectively. 
			
Motivated by the importance of PHS,  one of us has recently proposed an explicitly particle-hole symmetric theory of the FQH effect---the Dirac composite fermion theory~\cite{Son:2015xqa}. A distinctive feature of this theory is that, in gapless $\nu=\frac12$ state, PH conjugation maps a composite fermion to itself, only reversing the direction of its momentum (similar to time reversal for ordinary fermions).  With this assumption, the Dirac-ness of the composite fermion (its Berry phase $\pi$ around the Fermi circle) is an unavoidable consequence of the properties of the square of the particle-hole conjugation operator~\cite{LevinSon:unpublished,Geraedts197}. Numerical simulation provides the most nontrivial check for the this Berry phase~\cite{Geraedts197}, which is starting to be explored experimentally~\cite{Pan:2017}.
The Dirac composite fermion theory partially explains the experimentally measured disparity between the density of electrons and density of the composite fermions \cite{Kang:1993imw,Kamburov:2014,Cheung:2016}. 
The proposal of the Dirac composite fermion theory has stimulated the conjecture about a large web of field-theoretic dualities in 2+1 dimensions \cite{Metlitski:2015eka,Wang:2015qmt,Karch:2016sxi,Seiberg:2016gmd}.
	
In light of the above, an important question arises about whether or not the HLR theory is fundamentally inconsistent with particle-hole symmetry. As mentioned, early attempt \cite{Kivelson:1997} to check PHS within the HLR theory was unsuccessful. However, a recent reanalysis~\cite{Wang:HLR} finds that for some physical quantities, including the Hall conductivity in the presence of particle-hole symmetric disorder and the location of magnetoroton minima, the HLR theory gives particle-hole symmetric results that coincide with those of the Dirac composite fermion theory. 
The authors of Ref.~\cite{Wang:HLR} made a conjecture that the HLR theory has an emergent particle-hole symmetry in the infrared and it is indistinguishable from the Dirac composite fermion theory as far as physical observables are concerned. Re-examining this claim is another goal of this paper.

In this work, we compute various physical quantities for fractional quantum Hall states in the Jain sequences with filling factors $\n=\frac N{2N+1}$ and $\nu=\frac{N+1}{2N+1}$.  Treating $1/N$ as a small parameter, we develop an efficient method to compute correlation functions in the Dirac composite fermion theory.
We then compare with the HLR theory (in its phenomenologically most successful improved version, the MMRPA theory) and check for the presence of particle-hole symmetry. We find, unsurprisingly, that the results derived from the Dirac composite fermion theory satisfy the requirements of PHS. The situation with the HLR theory turns out to be quite intriguing. As we expect, the correlation functions computed from the HLR theory are not particle-hole symmetric. We also observe a violation of the Haldane bound on the leading $q^4$ coefficient of the projected static structure factor~\cite{Haldane:bound1,Haldane:bound2}.
Surprisingly, however, the dispersion relation of the neutral excitations is particle-hole symmetric (to leading and next-to-leading order in the large-$N$ expansion), and moreover coincides with the result obtained from the Dirac composite fermion theory by setting all Landau's parameters to zero.  Thus we conclude that the claim of Ref.~\cite{Wang:HLR} about emergent particle-hole symmetry of the HLR theory is invalid as far as current versions of the latter are concerned, but it is unclear if it can be made valid again by, say, additional improvements to the HLR theory on top of those already made in the MMRPA.


The layout of the paper is as follows. In Sec.~\ref{sec:review} we review the framework of the Dirac composite fermions. We derive the Lagrangian of the effective field theory describing FQH states with filling fraction given by Jain's sequence $\nu_{\pm}=\frac{1}{2}\pm \frac{1}{2(2N+1)}$ and emphasize its origins in particle-hole symmetry and Galilean invariance.  In Sec.~\ref{sec:semiclassical} we present the Fermi-liquid formalism which is the main computational framework of this paper. We derive a set of recursion relations and boundary conditions that enable us to compute response functions and dispersion relations in closed form.
In Sec.~\ref{sec:GMP} we derive from the Dirac composite fermion theory the long-wavelength limit of the celebrated Girvin--Macdonald--Platzman algebra and demonstrate the crucial role of the dipole moment of the composite fermions in this derivation.  We discuss the dispersion relation of the neutral excitations of the theory in Sec.~\ref{sec:dispersion} and compare with the results from HLR theory and the numerical work~\cite{Scarola:2000}. In Sec.~\ref{sec:response} we compute the susceptibility and the Hall conductivity and comment on their relations to various topological quantities. In both Sec.~\ref{sec:dispersion} and Sec.~\ref{sec:response}, we point out the expectations based on particle-hole symmetry and whether or not they are satisfied in the different theories under consideration. We conclude in Sec.~\ref{sec:conclusion}.  The Appendix contains additional technical details.

\section{Effective field theory of FQH near half filling}
\label{sec:review}

\subsection{Review of the Dirac composite fermion}

Let us begin with a heuristic overview of the composite Dirac fermion. We start in flat space first. Working to lowest order in the derivative expansion, the action, as proposed in Ref.~\cite{Son:2015xqa} is
\begin{equation}\label{eq:naive_Dirac}
	S(\psi,a,A)=\int \frac{i}{2}\bar{\psi}\g^0 \overset{\leftrightarrow}{D}_0 \psi + \frac{i}{2} v_F \bar\psi \g^i \overset{\leftrightarrow}{D}_i \psi - \frac{1}{4\pi} adA + \frac{1}{8\pi}AdA ,
\end{equation}
where $D_\mu = \d_\mu - i a_\mu$. For the Dirac matrices we choose the representation $\gamma^0=\sigma^3$, $\gamma^i=\sigma^3\sigma^i$, $i=1,2$.
Both the Dirac field and the gauge field $a$ are dynamical, while $A$ is an external background field. $v_F$ is a phenomenological parameter of Dirac composite fermion theory (replacing the effective mass in HLR theory). Consider for the moment a background constant magnetic field. Since $a$ appears linearly it simply acts as a Lagrange multiplier, enforcing a constant composite fermion density and a vanishing current,
\eq{
\psi^\dagger\psi  = \frac{B}{4\pi}  \rightarrow N_{CF} = \frac{N_\phi}{2},
  \qquad \bar\psi \g^i \psi = 0.
}
Note that (unlike in the usual flux attachment approach) the number of composite fermions $N_{CF}$ is always half the magnetic flux $N_\phi$, even away from half-filling. We can also calculate the charge density by taking $\delta S/\delta A_0$, 
\eq{
J^0 = \frac{B-b}{4\pi}\rightarrow N_e=\frac{N_\phi}{2}-\frac{n_\phi}{2}.}
In particular, we can move away from half filling by turning on a nonzero background $b$ \footnote{That means restricting the integration over a sector with a fixed total flux of $b$.}. 

One can work out the filling factors that correspond to composite fermions forming an integer quantum Hall state.
Recall that if we consider zero fermion number to be the zeroth Dirac Landau level half filled and that each Landau level has $|n_\phi|$ states, so if we fill all negative levels, the zeroth, and $N$ positive energy Landau levels the composite fermion number is $|n_\phi|(N+1/2)$, which implies $n_\phi = \pm \frac{N_\phi}{2N+1}$. From this we can directly calculate the filling fraction
\eq{
\nu = \frac{N_e}{N_\phi} = \frac 12\mp \frac{1}{4N+2} = 
\left\{\begin{array}{ccc}\frac{N}{2N+1} & : & n_\phi>0 \\\frac{N+1}{2N+1} & : & n_\phi<0\end{array}\right.
,
}
 yielding either the standard or conjugate Jain series.
 
 It is possible to convince oneself that the shift of these states comes out correctly as well.
 If we are in a curved background, we must account for the quantum Hall shift by coupling to background curvature. Following Ref.~\cite{Prabhu:2017rz}, this requires covariantizing our spinor derivative to $D_\mu = \d_\mu - i a_\mu + \frac{i}{2}\sigma^3 \omega_\mu$ (as the composite Dirac fermion is spin half) and shifting $A \rightarrow A + \frac 12 \omega$,
\begin{equation}\label{eq:SAaom}
S(\psi, a, \w, A) =
\int i \bar\psi \gamma^\mu D_\mu \psi - \frac{1}{4\pi} ad(A+\frac 12 \omega ) + \frac{1}{8\pi}(A+\frac 12 \omega ) d (A+\frac 12 \omega ).
\end{equation}
Consider now the composite fermion on a sphere. Due to coupling to curvature, the $N$th Dirac Landau level has $|n_\phi|+2N$ states, meaning we must generate two Chern-Simons terms when integrating out the fermions:
\begin{equation}\label{eq:adaadom}
\pm \frac{N+1/2}{4\pi}ada + \frac{N(N+1)}{4\pi}ad\omega,
\end{equation}
where the sign depends on the sign of $n_\phi$. It is now trivial to perform the Gaussian integral over $a$ to find the topological action
\eq{
\frac{N}{2N+1}\frac{1}{4\pi}AdA + \frac{N(N+2)}{2N+1}\frac{1}{4\pi}Ad\omega,~n_\phi>0,
}
\eq{
\frac{N+1}{2N+1}\frac{1}{4\pi}AdA + \frac{(N+1)(1-N)}{2N+1}\frac{1}{4\pi}Ad\omega,~n_\phi<0,
}
reproducing both the correct Hall conductance and topological shift for both Jain states found in Ref.~\cite{Nguyen:2016ph}.

\subsection{Further constraints }


As a candidate for the low-energy effective theory,
the Dirac action we have considered so far is incomplete for a few reasons. Since it is an effective field theory for electrons in the lowest Landau level it must inherit all of the symmetries of the lowest-Landau-level (LLL) problem, including Galilean symmetry. The problem of modifying the Dirac action to make it into a low-energy effective theory satisfying Galilean invariance can be solved by using the apparatus of Newton-Cartan geometry~\cite{Prabhu:2017rz}.  For completeness, we explain how to construct a Galilean theory of the Dirac composite fermion here, mostly without proof (for details see Ref.~\cite{Prabhu:2017rz})
First of all, we find that the time derivative term is not invariant under Galilean boosts as it is written (in the standard Newton-Cartan conventions)
\begin{equation}
\frac{v^\mu}{2}\psi^\dagger \overset{\leftrightarrow}{D}_\mu \psi,~v^\mu \rightarrow v^\mu + \delta v^\mu.
\end{equation}
However the existence of a strong magnetic field provides us with a preferred reference frame, the drift velocity
\begin{equation}\label{eq:drift_vel}
u^\mu = \frac{\ce^{\mu\nu\rho}F_{\nu\rho}}{2B} = \begin{pmatrix} 1 \\ E^y/B \\ -E^x/B \end{pmatrix}.
\end{equation}
We can therefore construct the time derivative with $u$:
\eq{
i\frac{u^\mu}{2}\psi^\dagger \overset{\leftrightarrow}{D}_\mu \psi = \frac{i}{2}\psi^\dagger \overset{\leftrightarrow}{D}_t \psi + i \frac{\ce^{ij}E_j}{2 B}\psi^\dagger \overset{\leftrightarrow}{D}_i \psi.
}
The second term on the right hand side can be interpreted as interaction energy of the electric field with dipoles~\cite{Prabhu:2017rz} $\mathbf E\cdot\mathbf d$, with the density of electric dipole moment $\mathbf d$ given by
\begin{equation}
  d_i = -\frac i{2B} \epsilon_{ij} \bar\psi \gamma^0 \overset{\leftrightarrow}{D}_j \psi \equiv \frac1 B \epsilon_{ij} T^{0j},
\end{equation}
where $T^{0j}$ is the momentum density carried by the composite fermion.  This is in contrast to 
the naive action~(\ref{eq:naive_Dirac}) where $\psi$ does not couple directly to the external gauge field.  One can say that each composite fermion quasiparticle with momentum $\mathbf p$ carries an dipole moment with respect to the external electric field 
orthogonal to the momentum,
\begin{equation}
\mathbf{d}=-\ell_B^2 \mathbf{p}\times \hat{\mathbf{z}}.
\end{equation}
The composite fermion dipole moment has been discussed in earlier works \cite{Read:1994,zhang1992}. 

There are additional constraints that come from inheriting the symmetries of the theory of massless fermion with $g$-factor equal to 2~\cite{Prabhu:2017rz}. First of all, the electromagnetic gauge field is shifted by a term proportional to the vorticity of the drift velocity,
\eq{
A \rightarrow A - \frac 12 (\nabla\times u )dt = A + \frac{\nabla\cdot E}{2B}dt,
} 
and  the spin connection the composite fermion couples to also gets a term proportional to this vorticity, such that even in flat space 
\begin{equation}\label{eq:omegaEB}
\omega = -\frac{\nabla \cdot E}{2B} dt.
\end{equation}
This means that even in flat space the Chern-Simons terms can be collected into one object
\begin{equation}\label{eq:cA}
\cA = A - \frac 12 (\nabla\times u) dt + \frac 12 \omega = A + \frac{\nabla\cdot E}{4B} dt,
\end{equation}
giving the flat space action (with no long range interactions)
\begin{equation}\label{eq:DCF2}
	S(\psi,a,A)=  \int i\psi^\dagger D_t \psi+ i v_F \psi^\dagger \sigma^i D_i \psi+ \frac{i}{2} u^i\psi^\dagger \overset{\leftrightarrow}{D}_i \psi  
	- \frac{1}{4\pi} ad \cA + \frac{1}{8\pi} \cA d \cA,
\end{equation}
where 
\begin{equation}\label{eq:Dmu}
D_\mu = \d_\mu - i a_\mu +\frac i2 \sigma^3 \omega_\mu\,.
\end{equation}

\subsection{Large-$N$ counting}

Before proceeding with further modifications, we pause here for a discussion of the large $N$ limit an the scaling of various quantities with $N$.  

We are interested in the electromagnetic response of a system with $\nu=\frac12+\cO(N^{-1})$, for energy and momentum of order $\cO(1/N)$.  Space and time derivatives thus count as $1/N$:
\begin{equation}
  \d_t \sim \d_x \sim \frac1N
\end{equation}
For example, the difference between $\cA$ and $A$ [Eq.~(\ref{eq:cA})] is relatively of order $1/N^2$.  One can view the term $\cA d\cA$ as containing terms of different powers of $N$,
\begin{equation}
  \cA d \cA \sim \frac{1}{N} A^2 + \frac{1}{N^3} A^2 + \cdots
\end{equation}
(we are interested only in terms quadratic in $A$). 
The linear response over external perturbations of $A_\mu$ is given by the polarization tensor of the system as $\Pi_{\mu\nu}$. 
The contact term $\cA d\cA$ thus contributes terms of order $1/N$ and $1/N^3$ to $\Pi_{\mu\nu}$.   For a rotationally invariant system, due to charge conservation, there are 3 independent components of $\Pi_{\mu\nu}$: the susceptibility $\chi\sim\Pi_{00}$, the Hall conductivity $\sigma^H\sim \omega^{-1}\Pi_{0\perp}$, and the transverse response function $\Pi_{\perp\perp}$, where $\perp$ denotes the spatial direction perpendicular to the momentum.  In this paper we will be interested in $\chi$ and $\sigma^H$. Due to the presence of the factor $\omega^{-1}\sim N$ in its definition, the Hall conductivity that arises from $\cA d\cA$ contains terms of order $1$ and $1/N^2$. In fact, it can be computed explicitly from Eq.~(\ref{eq:cA}) to be  
 \begin{equation}
  \sigma^H(\omega, q)|_{\cA d \cA} = \frac1{4\pi} \left ( 1- \frac{q^2}4 \right),  
\end{equation}
This up to the factor of $\frac12$ matches with the exact result for the Hall conductivity of a full Landau level~\cite{SonLevin:PH}. Thus, the $\frac1{8\pi}\cA d\cA$ term in Eq.~(\ref{eq:DCF2}) encodes one half of the response function of a full Landau level (the $\nu=1$ state).  The susceptibility $\chi$, formally of order $1/N$, turns out to be zero in the term $\cA d\cA$.  Physically, the full Landau level is completely inert to fluctuations of the scalar potential in the LLL limit.

We will compute it by first integrating over $\psi$ and then integrating over $a_\mu$ in the saddle-point approximation.  The saddle-point value of $a$ is $\sim A/N$ [see Eqs.~(\ref{eq:SAaom}) and (\ref{eq:adaadom})]. Thus, we can estimate
\begin{equation}
   a d \cA \sim \frac{A^2}{N^2}\,.
\end{equation}
By evaluating the term $ad\cA$ to next-to-leading order in $1/N$, we will have get terms up to $A^2/N^3$ inclusively. To that order one can replace $\cA$ by $A$ in the $ad\cA$ term, but not in the $\cA d\cA$ term.

The fermion couples, though various fermion bilinears, to $a_\mu$, $u^i$, and $\omega_\mu$.  Ignoring $\omega_\mu$ for now, integrating over the fermion one obtains schematically, to quadratic order
\begin{equation}\label{eq:a2auu2}
    a^2 + a u + u^2,
\end{equation}
where the coefficients are of order one. Since $a,u\sim A/N$, the fermion loops contribute $A^2/N^2$ to the partition function, comparable to the term $adA$. We will compute this loop to next-to-leading order in $1/N$. To that order, when calculating the drift velocity $u^i$, there is no difference if one uses in Eq.~(\ref{eq:drift_vel}) the improved gauge potential (\ref{eq:cA}) or the unimproved one.

Thus, evaluating the fermion loop up to the next-to-leading order, one will be able to find the $\cO(1)$, $\cO(1/N)$, and $\cO(1/N^2)$ terms in $\sigma^H$, and the $\cO(1/N^2)$ and $\cO(1/N^3)$ terms in the susceptibility. It may seem strange that a next-to-leading order calculation would give us the Hall conductivity with a precision of $1/N^2$; this is because the $\cO(1)$ term is completely trivial (equal to $1/4\pi$).

The spin connection $\omega$ is of order $A/N^2$, according to Eq.~(\ref{eq:omegaEB}), and is $1/N$ smaller than $a_\mu$.  Thus it may appear that its contribution will be only $1/N$ suppressed compared to the terms in Eq.~(\ref{eq:a2auu2}), and has to be taken into account in a next-to-leading-order calculation. However, the operator that $\omega$ couples to, $\psi^\dagger\sigma^3\psi$, has vanishing matrix elements between states near the Fermi surface, so the diagrams containing it are further suppressed by additional powers of $1/N$. Thus,  $\omega_\mu$ affects the fermionic loop only in the next-to-next-to-leading order in $1/N$, and can be safely ignored in our calculations.

\subsection{Coulomb interactions}

The one piece missing in our theory is the inclusion of long range Coulomb interactions. 
As discussed in Ref.~\cite{SonLevin:PH}, including long-range interactions for electrons requires modifying the effective field theory \eqref{eq:DCF2} as follows. First, we must obviously include a density-density interaction term
\begin{equation}\label{eq:Hartree}
 -\frac\alpha2 \! \int\! dt\, d^2\mathbf{x}\, d^2\mathbf{y}\, 
 \frac{\delta \rho(\mathbf{x})\delta \rho(\mathbf{y})}{|\mathbf{x}-\mathbf{y}|}.
\end{equation}
However, as explained in Ref.~\cite{SonLevin:PH}, this addition alone is not sufficient.  
One should also add contact terms, whose form is exactly fixed by particle-hole symmetry. 
According to the recipe of Ref.~ \cite{SonLevin:PH}, the effect of these
additional terms is to replace $\cA$ in all terms in Eq.~(\ref{eq:DCF2}) \emph{except for the $\cA d\cA$ term} by
 \eq{
\tilde A = \cA + \sum_{n}C_n \nabla^{2n}\delta B dt,
}
where $C_n$ are constants that can be determined from the electron two-body interaction potential. In some sense these terms constitute
the Fock (as opposed to Hartree) contribution to the self-consistent scalar potential acting on each electron  (for details see Ref.~\cite{SonLevin:PH}). This modification is not essential for the calculation of the susceptibility (in which only $A_0$ is perturbed), but important for that of the Hall conductivity.

Since $\delta B\sim A/N$, to the order we are working 
only the $C_0$ term contributes, and the action can be written as
\begin{multline}\label{eq:Lag0}
	S(\psi,a,A)=\int \frac{i}2\psi^\dagger \overset{\leftrightarrow}{D}_t \psi+ \frac i2 v_F\psi^\dagger \sigma^i \overset{\leftrightarrow}{D}_i \psi+ \frac{i}{2} u^i\psi^\dagger \overset{\leftrightarrow}{D}_i \psi 
	 -\frac{1}{4\pi}ad\tilde A + \frac{1}{8\pi} \cA d \cA \\ 
	 -\frac{\alpha}{2} \int\! dt\, d^2\mathbf{x}\, d^2\mathbf{y}\, \frac{\delta \rho(\mathbf x)\delta \rho(\mathbf y)}{|\mathbf x-\mathbf y|}.
\end{multline}
with
\begin{equation}
   \tilde A = A + C_0 \delta B dt, \qquad C_0 = \frac{\sqrt{\pi}\alpha}{4\sqrt{2}}\,,
\end{equation}
where we have given the value of $C_0$ for the Coulomb potential.
It is convenient to split various quantities into the average (or equilibrium) piece and a perturbation piece, e.g. the electron density is given by $\rho=\bar{\rho}+\delta \rho$ where the average and the perturbations of the density are given by:
\begin{align}
\bar{\rho}=\frac{B-b_0}{4\pi},\qquad
\delta \rho= \frac{1}{4\pi}  \vec{\nabla}\times \left( \delta \vec{A} -  \delta \vec{a} \right),
\end{align}
where $b_0=\pm \frac{B}{2N+1}$ is the effective magnetic field felt by the composite fermions. 
In what follows, we find it useful to rewrite the long range density-density Coulomb interaction interaction by introducing a Hubbard-Stratonovich transformation via the field $\phi$:
 \begin{multline}\label{eq:L-tedious}
   L=\int\! d^2 \mathbf{x}\,\Bigl(\frac i2 \bar\psi\gamma^0 \overset{\leftrightarrow}{D}_0\psi+	\frac{i v_F}2 \bar\psi\gamma^i \overset{\leftrightarrow}{D}_i\psi -\frac{1}{4\pi}ad\tilde{A}+\frac{1}{8\pi}\cA d \cA
 +\frac{i}{2}u^i\psi^\dagger \overset{\leftrightarrow}{D}_i \psi \Bigr) \\
  +\int d^2 \mathbf{x} \frac{1}{4\pi}\phi\,\vec{\nabla}\times\big(\delta \vec{A}- \delta \vec{a}  \big) +\frac{1}{4\pi\alpha}\int \frac{d^2 \mathbf{q}}{(2\pi)^2}\frac{\phi(-\mathbf{q})\phi(\mathbf{q})}{q} ,
 \end{multline}
where we have also performed a Fourier transform on the last term.   As $\phi$ is the Hartree contribution to the scalar potential, for the drift velocity in Eq.~(\ref{eq:L-tedious}) one should use
\begin{equation}
\label{eq:drift}
u_i=\frac{\epsilon^{ij}(\tilde E_j +\partial_j \phi)}{B}\,,
\end{equation}
where $\tilde E_j=\d_j \tilde A_0 -\d_0 A_j$.

In principle, the Lagrangian~(\ref{eq:L-tedious}) can be used for calculation. One can, for example, develop an RPA by integrating over $\psi$ in the Fermi liquid ground state of the latter, keep only the quadratic terms in the result.
This requires the calculation of the fermion loop diagrams involving insertions of the current operator $\bar\psi\gamma^\mu\psi$ or the momentum density $\bar\psi\gamma^i \overset{\leftrightarrow}{D}_i\psi$. 

\subsection{A little trick}

If one is interested in the regime of small energy and momentum, e.g., when these are suppressed by a factor of $1/N$ with $N$ being a large number, then 
one can employ a useful mathematical trick developed in Ref.~\cite{NS}. 
One can show that the following two fermionic theories are equivalent at large $N$ for \emph{any} $u^i$:
\begin{equation}\label{eq:dirac_N_equiv}
\frac{i}{2} \bar{\psi}\gamma^0 \overset{\leftrightarrow}{D}_0\psi+	\frac{i v_F}{2} \bar{\psi}\gamma^i \overset{\leftrightarrow}{D}_i\psi
+\frac{i}{2}u^i\psi^\dagger \overset{\leftrightarrow}{D}_i \psi \approx \frac{i}{2} \bar{\psi}\gamma^0 \overset{\leftrightarrow}{\tilde{D}}_0\psi+	\frac{i v_F}{2} \bar{\psi}\gamma^i \overset{\leftrightarrow}{\tilde{D}}_i\psi,
\end{equation}
where on the right hand side
\begin{equation}\label{eq:tilde_a}
\tilde{D}_\mu=\partial_\mu-i  \tilde{a}_\mu,\qquad
\tilde{a}_0=a_0 + \frac{m_*}2 u_i u^i, \qquad \tilde{a}_i=a_i - m_* u_i\,,
\end{equation}
where $m_*=k_F/v_F$ is the effective mass.  The sign ``$\approx$''in Eq.~(\ref{eq:dirac_N_equiv}) means that if one integrates over the fermion, the resulting functionals of $a_\mu$ and $u_i$ coincide to leading and next-to-leading order in $1/N$ (provided the energy and momentum scales are of order $1/N$).

In Ref.~\cite{NS} this equivalent is checked by direct calculation. The physical basis for  the equivalence~(\ref{eq:dirac_N_equiv}) is the proportionality between the momentum density and the current density: for quasiparticles near the Fermi surface, the former is $m_*$ times the latter.  This fact allows us to absorb the $u$-coupling to $\psi$ into the gauge field $a$.

Using this equivalence, we can rewrite the Lagrangian (\ref{eq:Lag0}) as 
\begin{multline}
\label{eq:after_duality}
L =\int\! d^2 \mathbf{x}\, \Bigl(\frac{i}{2} \bar{\psi}\gamma^0 \overset{\leftrightarrow}{\tilde{D}}_0\psi+	\frac{i v_F}{2} \bar{\psi}\gamma^i \overset{\leftrightarrow}{\tilde{D}}_i\psi-\frac{1}{4\pi}a d\tilde{A}+\frac{1}{8\pi}\cA d \cA\Bigr) \\
  -\int\! d^2 \mathbf{x} \frac{1}{4\pi}\phi \vec{\nabla}\times \bigl( \delta \vec{a} - \delta \vec{A}\,\bigr)+\frac{1}{4\pi\alpha}\int\! \frac{d^2 \mathbf{q}}{(2\pi)^2}\,\frac{\phi(-\mathbf{q})\phi(\mathbf{q})}{q} \,.
\end{multline}

One by-product of \eqref{eq:dirac_N_equiv} is that the composite fermion is not directly coupled to $\phi$.
 Integrating $\phi$ out and redefining $\tilde{a}_\mu \rightarrow a_\mu$ for brevity, we arrive at the final form of the Lagrangian: 
\begin{multline}
\label{eq:L_final}
L = L_{CF}
+ \!\int\! d^2 \mathbf{x}\,  \Bigl(-\frac{1}{4\pi}a d\tilde{A}+\frac{1}{8\pi}\cA d \cA-\frac{m_*}{8\pi B}\tilde{E}_i\tilde{E}^i\Bigr)\\
 -\frac{\alpha}{16\pi} \!\int\! \frac{d^2 \mathbf{q}}{(2\pi)^2} \, \frac{q_iq_k\left[\epsilon^{ij}\left(\delta a_j-\delta A_j\right)-\frac{m_*}{ B}\tilde{E}^i\right]\!(-\mathbf{q})\left[\epsilon^{kl}\left(\delta a_l-\delta A_l\right)-\frac{m_*}{B}\tilde{E}^k\right]\!(\mathbf{q})}{q-\frac{m_*}{2  B}\alpha q^2}\,, 
\end{multline}
where
\begin{equation}\label{eq:finalL}
L_{CF}=\int \!d^2 \mathbf{x}\, \frac{i}{2}  \Big(\bar{\psi}\gamma^0 \overset{\leftrightarrow}{{D}}_0\psi+	v_F \bar{\psi}\gamma^i \overset{\leftrightarrow}{{D}}_i\psi\Big), \qquad
{D}_\mu=\partial_\mu-ia_\mu \,.
\end{equation}
For convenience, here we recall our notations for various modified gauge potentials and gauge fields appearing in Eq.~(\ref{eq:L_final}):
\begin{equation}
  \cA = A + \frac{\nabla\cdot E}{4B} dt, \qquad
  \tilde A = A + \frac{\sqrt\pi \alpha}{4\sqrt2}\delta B\, dt, \qquad
  \tilde E_i = \d_i \tilde A_0 - \d_0 A_i \,.
\end{equation}


\section{Semiclassical approximation}
\label{sec:semiclassical}
%

In this Section we develop a semiclassical approximation which is the main calculation tool of this paper. This allows one to effectively carry out the integration over the fermions. The goal is to analyze Jain's states at fillings  $\nu=\frac{N}{2N+1},\frac{N+1}{2N+1}$ where the Dirac composite fermion forms integer quantum Hall states with filling fraction $\nu_{CF}=\pm (N+\frac{1}{2})$. 

Since we are interested in states near half filling $\nu \sim \frac12$, we will take $N$ to be large and use $1/N$ as an expansion parameter. Here the composite fermion lives in an average magnetic field of order $1/N$.  The cyclotron frequency of the composite fermion (the parameter $\omega_b$ introduced below) goes to zero in the large $N$ limit (in the case of Coulomb interaction as $1/N$ up to a logarithm), and the radius of the semiclassical orbit of a composite fermion diverges as $N$. The regime of nontrivial physics is $\omega\sim\omega_b$, $q\sim 1/R=\cO(N^{-1})$. 

We reiterate that in this limit, one can show that the RPA and semiclassical calculations are equivalent up to leading and next-to-leading orders in $1/N$ expansion \cite{nguyenemIQH}. However, in what follows we will focus on the semiclassical formalism which would allow us to derive closed form results. It will also allow us to generalize the Dirac composite fermion theory by introducing short range interactions through Landau parameters of the Fermi-liquid model. Note that even though this generalization can also be incorporated in the RPA approximation, the implementation would be prohibitively complicated.

	\subsection{Quantization of the Fermi surface fluctuations}		\label{sec:Fermi_Liquid_from_commutations}


In this method we look at the fluctuations of the shape of the Fermi surface, bosonize and study the commutation algebra governing these fluctuations.  This
procedure was studied 
previously by by Haldane~\cite{Haldane:1994} (see also Refs.~\cite{Houghton:1992dz,CastroNeto:1994,Mross:2011}).
Here we recall a simple and intuitive semiclassical derivation of this algebra motivated by considering Poisson brackets of operators in magnetic fields (for details see Ref.~\cite{GNRS1}). In the next subsection we will rederive this algebra using the quantum Boltzmann's equation.

\begin{figure}
	\centering
	\includegraphics[width=16em]{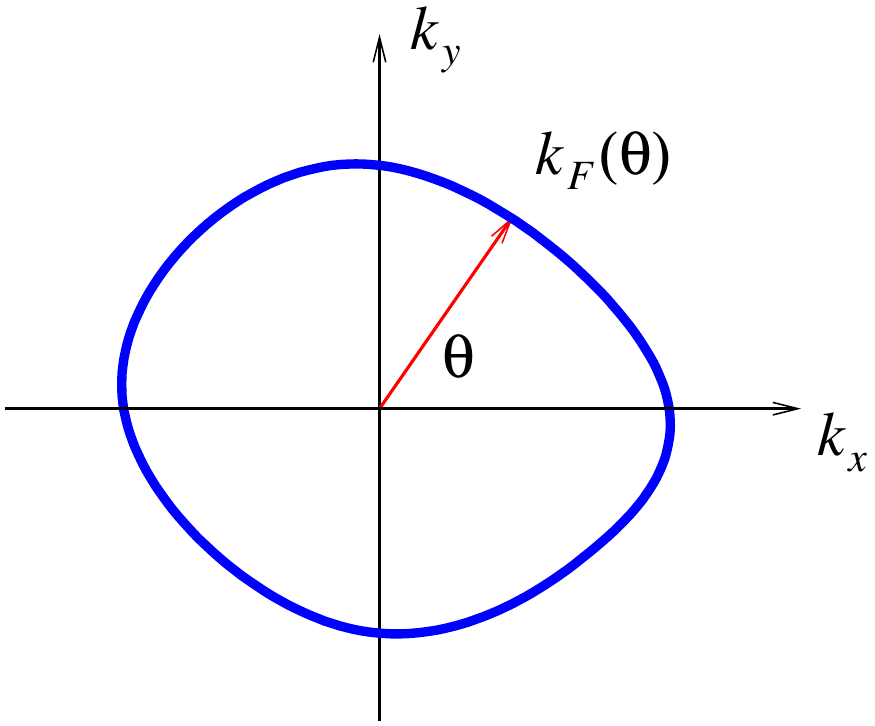}
	\caption{A deformed Fermi surface.}
	\label{fig:fs}
\end{figure}

We assume that low-energy, long-wavelength excitations of a Fermi liquid can be described by fluctuations of the shape of the Fermi surface (Fig.~\ref{fig:fs}), parametrized by an infinite number of fields $u(\theta)$ or $u_n$,
\begin{equation}
k_F(t,{\bf x},\theta) = k_F^0+u(t,{\bf x},\theta)
= k_F^0 + \!\!\! \sum_{n=-\infty}^\infty\!
u_n(t,{\bf x})\, e^{in\theta}.
\end{equation}
The quasiparticle distribution
function $n_{\bf p}(t,{\bf x})$ is one inside the Fermi line and zero
outside the line: for $\mathbf p =(p\cos\theta,\,p\sin\theta)$,
\begin{equation}
  n_{\mathbf p}(t,\mathbf x) = \theta(k_F(t,\mathbf x,\theta)-p),
\end{equation}
where $\theta$ is the step function.

The commutation
relation between the $u_n$'s can be determined from a semiclassical matching calculation.
For each function on the phase space, $F(\mathbf{x},\, \mathbf{p})$, one defined an operator $F$
\begin{equation}
F = \int\! \frac{d^2{\bf x}\,d^2{\bf p}}{(2\pi)^2}\, F({\bf x},{\bf p}) n_{\bf p}
({\bf x}),
\end{equation}
where $n_{\bf p}({\bf x})$ is the quasiparticle distribution function, defined above. Obviously $F$ is a functional of $u_n$.  To linear order in $u_n$, 
\begin{equation}
  F = \int\!d^2\mathbf x \! \int\limits_{|\mathbf p|<k_F} \! \frac{d^2\mathbf p}{(2\pi)^2}\, F(\mathbf x,\, \mathbf p)
  + k_F\! \int\!d^2\mathbf x\! \int\limits_0^{2\pi}\!d\theta\, F(\mathbf x,\, k_F\mathbf n_\theta) u(\mathbf x,\, \theta)
\end{equation}
with $\mathbf n_\theta = (\cos\theta,\, \sin\theta)$.

Let us take two operators, $F$ and $G$, corresponding to two phase-space functions $F(\mathbf x\,,\mathbf p)$ and $G(\mathbf x\,,\mathbf p)$.  We impose the following condition on the commutation relation between $F$ and $G$, 
\begin{equation}\label{FG-commut}
[F,\, G] = -i \!\int\! \frac{d^2{\bf x}\,d^2{\bf p}}{(2\pi)^2}\,
\{F,\, G\} ({\bf x}, {\bf p}) n_{\bf p}({\bf x}),
\end{equation}
where the $\{F,\,G\}$ is the classical Poisson bracket between $F$ and $G$,
\begin{equation}\label{eq:FG}
\{F,\, G\} = \frac{\partial F}{ \partial p_i} \frac{\partial G}{\partial x_i}
- \frac{\partial G}{\partial p_i} \frac{\partial F}{\partial x_i}
- b\epsilon^{ij} \frac{\partial F}{\partial p_i}\frac{\partial G}{ \partial p_j} ,
\end{equation}
where we have allowed the the existence an external
magnetic field $b$ much smaller than the scale set by the Fermi momentum ($b\ll k_F^2$). Expanding both sides of Eq.~(\ref{eq:FG}) in series over $u$, one can determine the commutation relation for $u$. For example, the leading constant term in the commutator of $u$ follows from
\begin{multline}
  k_F^2 \!\int\!d^2\mathbf x\, d^2\mathbf x' \! \int\!d\theta\,d\theta'\, F(\mathbf x,\, k_F\mathbf n_\theta) G(\mathbf x, k_F\mathbf n_{\theta'})
  [u(\mathbf x\,, \theta),\, u(\mathbf x',\, \theta')]\\
  = -i \int\!d^2\mathbf x \! \int\limits_{|\mathbf p|<k_F} \! \frac{d^2\mathbf p}{(2\pi)^2}\, \{F,\, G\}(\mathbf x,\, \mathbf p)
\end{multline} 
Using Stokes' theorem the right hand side can be transformed to an integral over the the boundary of the Fermi disk.  At the end, we find
\begin{equation}
[u({\bf x},\theta),\, u({\bf x}',\theta')] =
\frac{i(2\pi)^2}{k_F} \biggl[
-n_i(\theta)\frac{\partial}{{\partial x_i}}
+ \frac{b}{k_F}\frac{\partial}{\partial\theta}
\biggr]  [\delta({\bf x}-{\bf x}')\delta(\theta-\theta')]
+O(u),
\end{equation}
 In terms of $u_n$, the
formula reads   
\begin{equation}\label{commutator}
[u_m({\bf q}),\, u_n({\bf q}')] =
\frac{2\pi}{k_F}\biggl[-\frac{b m}{k_F} \delta_{m+n,0}
+  \delta_{m+n,-1} q_{\bar{z}}
+ \delta_{m+n,1} q_{z}
\biggr]  (2\pi)^2\delta({\bf q}+{\bf q}') + O(u),
\end{equation}
where $q_z=\frac{1}{2}(q_1-iq_2),
q_{\bar{z}}=\frac{1}{2}(q_1+iq_2)$.  Note that the algebra depends only on the
size of the Fermi surface $k_F$, but not on any dynamic properties
(Fermi velocity, Landau's parameters, etc.).  These only enter the Hamiltonian.
Assuming the composite
fermions form a Fermi liquid, the
Hamiltonian of the system is
\begin{equation}\label{Hamiltonian}
H = \frac{v_F k_F}{4\pi}\!\int\!d^2{\bf x}\!\sum_{n=-\infty}^\infty\!
(1+F_n)u_n({\bf x}) u_{-n}({\bf x}),
\end{equation}
where $F_n$ are the Landau parameters.  

Within the theory (\ref{Hamiltonian}) one can rescale $v_F$ and all $1+F_n$ simultaneously so that the products $v_F(1+F_n)$ remain constants without changing any physics. Note that there is no relation between the effective mass $m_*=k_F/v_F$ and the Landau parameter $F_1$: Galilean invariance is enforced by the dipole electric moment of the composite fermion.

For Coulomb interactions, the composite fermions form a marginal Fermi liquid, and one should take $F_n$ to be the Landau parameters evaluated at the scale of the energy gap. For a Jain state at large $N$, these Landau parameters are expected to be proportional to $\ln N$ with the same prefactor. In this case, all the logarithms can be absorbed into the effective mass $m_*$, and all correlation functions are free from $\ln N$ divergences when expressed in terms of the new $m_*$. This is consistent with the cancellation of infrared divergences in electromagnetic response functions found in Ref.~\cite{Kim:1994}. We will not consider $\ln N$ to be a large parameter and will keep $F_n$ as free parameters in our further discussion.


The Hamiltonian~(\ref{Hamiltonian}) and the commutation
relations~(\ref{commutator}) form our theory of the neutral
excitations in the fractional quantum Hall fluid.  This theory involves
an infinite number of fields $u_n$, reminiscent of higher-spin
relativistic field theories~\cite{Fradkin:1987ks,Fradkin:1986qy}. Using commutator (\ref{commutator}), we obtain the linearized equation of motion for $u_n(\omega,\mathbf{q})$:
\begin{align}\label{EOMF}
[\omega + \sign(b_0)n(1+F_n)\omega_b] u_n =  v_F[q_z (1+F_{n-1}) u_{n-1}
+ q_{\bar{z}}(1+F_{n+1}) u_{n+1}] ,
\end{align}
where $\omega_b$ is the cyclotron frequency of the composite fermion,
\begin{equation}
\omega_b=\frac{|b_0|}{m_*} \,.
\end{equation}
We ignore nonlinear terms like $u_n\delta b $ in Eq.~(\ref{EOMF}).

\subsection{Derivation from quantum Boltzmann equation}		

We now repeat the derivation of the commutation relation, but this time from the perspective of the quantum Boltzmann equation.  This will simplify considerations of boundary conditions and as an added bonus we will be able to look at the response of the system to external fields. The derivation follows closely to the bosonization of Fermi liquid \cite{CastroNeto:1994,Haldane:1994,Polchinski:1992ed,Shankar:1993pf}.

Let us  again consider a fractional quantum Hall system in the Jain's sequence with filling fraction $\nu=\frac{N}{2N+1},\frac{N+1}{2N+1}$, which corresponds to a composite fermion with finite density $\bar{\rho}_{CF}=\frac{k_F^2}{4\pi}=\frac{B}{4\pi}$ in background magnetic field $b_0 = \pm B / (2N+1)$. In the large $N$ limit, i.e. $\nu \sim \frac12$, the effective magnetic field is small ($b_0 \sim k_F^2 / N$), allowing us to describe the system as a Fermi liquid with small deformations \cite{nguyenemIQH}. In other words, the composite Dirac fermions form a two-dimensional Fermi-surface with radius $k_F$. 

We now take the system to be in an effective magnetic and electric given by  $b(x,t)$ with average value $b_0$ and perturbation $\delta b=\epsilon^{ij}\partial_i \delta a_j$ and electric field $e_i=-\partial_0 \delta a_i$, where we have adopted the temporal gauge $a_0 = 0$. We will assume that $\delta b(x,t)$ and $\vec e(x,t)$ are weak and slowly varying ($\sim \mathcal O(1/N)$). 

The low energy physics of the Fermi liquid is described by a distribution function,
\begin{equation}
n_{\bf k}(t,{\bf x})=n^0(\mathbf{k})+\delta n_{\bf k}(t,{\bf x}) ,
\end{equation}	
with $n^0$ being the equilibrium fermionic distribution function
\begin{equation}
n^0(\mathbf{k})=\Theta(k_F-k) .
\end{equation}
The quantum Boltzmann equation in the collisionless limit then describes the evolution of perturbations  $\delta n_{\bf k}(t,{\bf x})$ 
\begin{multline}
\frac{\partial \delta n_{\bf k}(t,{\bf x})}{\partial t}+ \vec{v}(\mathbf{k})\cdot\vec{\nabla}_\mathbf{x}\delta n_{\bf k}(t,{\bf x}) +\vec{e}(\mathbf{x})\cdot\vec{\nabla}_\mathbf{k} \delta n_{\bf k}(t,{\bf x}) +(\vec{v}(\mathbf{k})\times \vec{b}(\mathbf{x}))\cdot\vec{\nabla}_\mathbf{k} \delta n_{\bf k}(t,{\bf x})\\ +\vec{v}(\mathbf{k})\cdot\vec{e}(\mathbf{x})\frac{\partial n^0(\mathbf{k})}{\partial \epsilon_\mathbf{k}}=0,
\end{multline}	
 where $\vec{v}(\mathbf{k})=\vec{\nabla}_\mathbf{k}\epsilon_\mathbf{k}$ is the group velocity derived from the dispersion relation $\epsilon_\mathbf{k}$. Since we are interested in the regime of
 frequency and momentum which are close to the Fermi surface, we can substitute
\begin{equation}
\delta n_{\bf k}(t,{\bf x})=u(\theta,\mathbf{x},t)\delta(k_F-k),
\end{equation}		
where $\theta$ is the direction of $\mathbf{k}$ on the Fermi surface (figure \ref{fig:fs}). Linearizing, we can rewrite the Boltzmann equation in terms of $u(\theta,\mathbf{x},t)$		
\begin{equation}
\label{eq:u1}
\frac{\partial u(\theta,\mathbf{x},t)}{\partial t}+ v_F \vec{n}_\theta\cdot\vec{\nabla}_\mathbf{x}u(\theta,\mathbf{x},t)  -\sign(b_0)\omega_b \frac{\partial u(\theta,\mathbf{x},t)}{\partial \theta}-\vec{n}_\theta\cdot\vec{e}(\mathbf{x})=0,
\end{equation}		
where $\vec{n}_\theta$ is defined as a unit vector normal to the Fermi surface at angle $\theta$,  the cyclotron frequency $\omega_b$ is given by  $\omega_b=\frac{|b_0|}{k_F/v_F}$ and we have ignored the higher order terms $\vec{e}\cdot\vec{\nabla}_\mathbf{k} \delta n_{\bf k}(t,{\bf x})$ and $(\vec{v}(\mathbf{k})\times \delta\vec{b}(\mathbf{x}))\cdot\vec{\nabla}_\mathbf{k} \delta n_{\bf k}(t,{\bf x})$. 

Performing a Fourier transform
\begin{equation}
u(\theta,\mathbf{x},t)=\int \frac{d^3q}{(2\pi)^3}u(\theta,\mathbf{q},\omega)e^{iq_\mu x^\mu} ,
\end{equation}		
Equation~\eqref{eq:u1} turns into
\begin{align}
-i\omega u(\theta,\mathbf{q},\omega)+iv_F \vec{n}_\theta\cdot\vec{q}u(\theta,\mathbf{q},\omega) -\sign(b_0)\omega_b \frac{\partial u(\theta,\mathbf{q},\omega)}{\partial \theta}-\vec{n}_\theta\cdot\vec{e}(\mathbf{q},\omega)=0.
\end{align}	
Similar to before we perform a mode decomposition as
\begin{equation}
u(\theta,\mathbf{q},\omega)=\sum_{n=-\infty}^{\infty}u_n(\mathbf{q},\omega)e^{i n \theta},
\end{equation}
which gives us the final equation of motion in the form of a recursion relation relating $u_{n+1}$ to $u_n = u_{n-1}$:
\begin{equation}
\label{eq:EOMF1}
0=(\omega + \sign(b_0) n \omega_b)u_n - \frac{B}{|b_0|}\omega_b(\tilde{q}_{\bar{z}}u_{n+1} 
 + \tilde{q}_z u_{n-1})+ \omega(\delta_{n,1}\delta a_z+\delta_{n,-1} \delta a_{\bar{z}}),
\end{equation}
where we use the short form $u_n$ for $u_n(\tilde{q},\omega)$ and we have defined 
\begin{align}
\ell_B=\frac{1}{\sqrt{B}}, \;\;
\delta a_z=\frac{1}{2}(\delta a_1-i\delta a_2),\notag\\
\tilde{q_i}=q_i \ell_B,\;\;
\delta a_{\bar{z}}=\frac{1}{2}(\delta a_1+i\delta a_2).
\end{align}

This equation of motion is the same as the system described by Hamiltonian (\ref{Hamiltonian}) and commutation relations (\ref{commutator}) when the Landau parameters are zero. Turning these on, the recursion relation becomes \cite{GNRS1} :
\begin{multline}
\label{eq:recur1}
0=\Big[\omega+\sign(b_0) n(1+F_n)\omega_b\Big]u_n
-\frac{B}{|b_0|}\omega_b\Big[\tilde{q}_{\bar{z}}(1+F_{n+1})u_{n+1}+\tilde{q}_z (1+F_{n-1})u_{n-1}\Big]\\
+\omega \big(\delta_{n,1}\delta a_z+\delta_{n,-1}\delta a_{\bar{z}}\big).
\end{multline}	
Equation  (\ref{EOMF}) is nothing but Eq.~(\ref{eq:recur1}) without $\delta a_i$, i.e., without the fluctuations of the background field. 

Finally, with the mode decomposition the composite fermion density $\rho_{CF}$ and current $J_{CF}^i$ can be rewritten as 
\begin{align}
\label{eq:unJ}
\rho_{CF}=&\int \frac{d^2 \mathbf{k}}{(2\pi)^2}n_{\bf k}(t,{\bf x}) =\bar{\rho}_{CF}+\frac{k_F}{2\pi}u_0, \notag\\ 
J^i_{CF}(\mathbf{x},t)=\int &\frac{d^2 \mathbf{k}}{(2\pi)^2}(1+F_1)n_{\bf k}(t,{\bf x})v^i(\mathbf{k})=\int \frac{d^2 \mathbf{k}}{(2\pi)^2}(1+F_1)n_{\bf k}(t,{\bf x})\frac{k^i}{m^*}\,,\notag\\
J_{CF}^1=\frac{k_F^2(1+F_1)}{4\pi m^*}&(u_1+u_{-1}), \qquad J_{CF}^2=\frac{ik_F^2(1+F_1)}{4\pi m^*}(u_1-u_{-1}),
\end{align}
and the continuity equation $\partial_0 \rho_{CF}+\partial_i J^i_{CF}=0$ turns into Eq.~\eqref{eq:recur1} for $n=0$.

\subsection{Boundary Conditions and Electromagnetic Responses}

So far in our discussion we have not included the electromagnetic field. In order to analyze the electromagnetic responses of the system as well as to  arrive at a closed set of equations we go back to the Lagrangian \eqref{eq:L_final}, we need to emphasize that the effective mass $m_*$ needs to be replaced  by $m_*/(1+F_1)$ in the appearance of Landau parameters. We consider the equations of motion for $a_\mu$.
The equations of motion for $a_0$ gives us the familiar constraint of composite fermion density 
\begin{equation}
\label{eq:constr1}
\rho_{CF}=\frac{B+\vec{\nabla}\times \delta \vec{A}}{4\pi}\,,
\end{equation}
while the equations of motion for $a_i$ in momentum space is
\begin{align}
\label{eq:constr2}
J_{CF}^i(q)=\frac{\epsilon^{ij }\tilde{E}_j(q)}{4\pi}+\frac{\alpha}{8\pi}\epsilon^{ij}q_j\frac{\left[\epsilon^{kl}q_k\left(\delta a_l(\mathbf{q})-\delta A_l(\mathbf{q})\right)-\frac{m_*}{(1+F_1) B}q_k\tilde{E}^k(\mathbf{q})\right]}{q-\frac{m_*}{2 (1+F_1) B}\alpha q^2}.
\end{align}
With the help of Eq.~(\ref{eq:unJ}), we can rewrite these constraint equations as 
\begin{align}
\frac{1{+}F_1}{2\pi}\frac{B}{m_*} u_1= &\, i\frac{\tilde{E}_z}{2\pi}-\frac{i\alpha}{4\pi(1-\frac{m_*}{2(1+F_1)B }\alpha q)}\frac{q_z}q\left[\epsilon^{ij}q_i \delta a_j-\epsilon^{ij}q_i \delta A_j-
\frac{m_*}{B(1{+}F_1)}q_i\tilde{E}^i\right],\notag\\
\label{eq:constr_k}
\frac{1{+}F_1}{2\pi}\frac{B}{m_*}u_{-1}=&-i\frac{\tilde{E}_{\bar {z}}}{2\pi}
+\frac{i\alpha}{4\pi(1-\frac{m_*}{2(1+F_1)B }\alpha q)}
\frac{q_{\bar{z}}}q \left[\epsilon^{ij}q_i \delta a_j-\epsilon^{ij}q_i \delta A_j-\frac{m_*}{B(1{+}F_1)}q_i\tilde{E}^i\right].
\end{align}
These relations as well as the continuity equation can be considered as boundary conditions  which along with the equation of motion \eqref{eq:recur1} provide us a closed set of equations which can be solved. Note that the continuity equation is automatically satisfied by the Bianchi identity and does not provide an independent constraint.

Finally, we can also derive the electromagnetic current in the composite Dirac fermion theory from the action via $J^\mu=\delta S/\delta A_\mu$ giving 
\begin{subequations}\label{eq:J}
\begin{align}
\rho=&\bar{\rho}+\frac{1}{4\pi} \Bigl(1+\frac{1}{4B}\partial_i\partial^i\Bigr)(\vec{\nabla}\times \delta \vec{A}-\vec{\nabla}\times \delta \vec{a})-\Bigl(1+\frac{1}{4B}\partial_i\partial^i\Bigr)\vec{\nabla}\cdot \vec{d},\\
J^i=&J_{CF}^i-\frac{\epsilon^{ij}e_j}{4\pi}+\frac{1}{16\pi B}\partial_i\partial^i(\vec{\nabla}\times \delta \vec{A}-\vec{\nabla}\times \delta \vec{a})+\frac{C_0}{4\pi}\epsilon^{ij}\partial_j(\vec{\nabla}\times \delta \vec{A}+\vec{\nabla}\times \delta \vec{a})\nonumber\\
&\hspace{150pt}+\partial_0 d_i+\frac{1}{4B}\partial_0\partial_i \vec{\nabla}\cdot \vec{d}-C_0\epsilon^{ij}\partial_j \vec{\nabla}\cdot \vec{d},
\end{align}
\end{subequations}
where $d_i$ is the dipole moment of composite fermion given by
\begin{equation}
\label{eq:dipole}
d_i=\frac{m_*}{(1+F_1)}\frac{\epsilon^{ij}}{B}J_{CF}^j\,.
\end{equation}
This completes the derivation of our semiclassical framework. The dynamics of the composite Fermi-liquid theory is given by the recursion relation~\eqref{eq:recur1} and  constraint equations \eqref{eq:constr1} and \eqref{eq:constr_k}. The electromagnetic responses of the theory can then be read off from Eq.~\eqref{eq:J}

\section{The Girvin--Macdonald--Platzman algebra}
\label{sec:GMP}

 The Girvin--Macdonald--Platzman (GMP) algebra generally refers to the commutation algebra governing the density operator of a fractional quantum Hall system projected to the lowest Landau level \cite{Girvin:1986zz}. Because of the projection, one can show that the resulting projected density operators are no longer commuting and their commutator is proportional to the projected density operator. In momentum space, the algebra takes the form:
  \begin{equation}
  [\rho(\mathbf{k}),\rho(\mathbf{k}')]\sim\sin(\mathbf{k}\times \mathbf{k}')
  		 \rho(\mathbf{k}+\mathbf{k}').
  \end{equation}		
  
The GMP commutation relation also appears in other contexts. It has been derived under the guise of $W_\infty$ algebra in Ref.~\cite{CAPPELLI:W}, where it was interpreted as the dynamical symmetry of area-preserving diffeomorphisms.  It also plays a role in the noncommutative field theory description of fractional quantum Hall system \cite{ISO:GMP,Susskind:NC,FRADKIN:NC} and in the Hamiltonian description of 
 the FQHE~\cite{Murthy:2003zz}.
  
The Dirac composite fermion theory is, by construction, an effective theory for interacting electrons projected to the lowest Landau level. In this section we show that we can reproduce the algebra, at leading order in the momentum expansion, directly from the Dirac composite fermion theory. As we will see, the fact that the composite fermion has electric dipole moment plays a crucial role in the appearance of the GMP algebra.  

 In what follows, we assume a constant background magnetic field $\delta A_i=0$. Starting from Eq.~\eqref{eq:J} and keeping only the leading and next-to-leading terms in the $1/N$ expansion, we can rewrite the charge density operator as  
\begin{eqnarray}\label{eq:rho-d}
\rho(\mathbf x)=\frac{B-b(\mathbf x)}{4\pi}-\partial_id_i(\mathbf x),
\end{eqnarray}
where $d_i$ is the electric dipole moment density of the composite Dirac fermion defined in (\ref{eq:dipole}). We also know that $d_i$ is related to composite fermion momentum density $T^{0i}$ by $d_i=B^{-1}\epsilon_{ij}T^{0j}$.  Using the canonical anticommutation relation for $\psi$ and $\psi^\dagger$, one can easily derive the commutator of $T^{0i}=-\frac i2 \psi^\dagger\overset{\leftrightarrow}{D}_i\psi$,
\begin{equation}
  [ T^{0i}(\mathbf x),\,  T^{0j}(\mathbf y)] =
  i \frac\d{\d y_i} ( T^{0j}\delta(\mathbf x-\mathbf y))
  - i \frac\d{\d x_j} ( T^{0i}\delta(\mathbf x-\mathbf y))
    + i \epsilon^{ij} b(\mathbf x)\psi^\dagger(\mathbf x)\psi(\mathbf x)\delta(\mathbf x-\mathbf y).
\end{equation}
Using this equation and Eq.~(\ref{eq:rho-d}), one finds
\begin{equation}\label{eq:leading-GMP}
[\rho(\mathbf{k}),\rho(\mathbf{k}')]=\frac{i\ell_B^2}{2\pi}(\mathbf{k}\times \mathbf{k}') \rho(\mathbf{k}+\mathbf{k}'),
\end{equation}
where we have put in the density of the composite fermions (\ref{eq:constr1}).  This is the long-wavelength limit of the GMP algebra.

Another, semiclassical, approach to the GMP algebra works as follows.
Recall also from Eq.~\eqref{eq:unJ} that the composite fermion current can be written in term of the semiclassical operator $n_{\vec p}(\vec y)$ as:
\begin{equation}
 J^i_{CF}(\mathbf{x})=\int\! { d^2 \mathbf{y}\, \frac{d^2 \mathbf{p}} {4\pi^2} \, \delta(\mathbf{x}-\mathbf{y})\frac{p^i(1+F_1)}{m_*}}n_{\mathbf{p}}(\mathbf{y}).
\end{equation}		
Combining these equations we have:
\begin{equation}
\rho(\mathbf x)=\frac{B-b(\mathbf x)}{4\pi}-\frac{1}{B} \!\int\! { d^2 \mathbf{y}\, \frac{d^2 \mathbf{p}}{4\pi^2}\,\epsilon^{ij}\frac{\partial}{\partial x^i}\delta(\mathbf{x}-\mathbf{y})p^j n_{\mathbf{p}}(\mathbf{y})}.
\end{equation}		
Noting the resemblance of the second term to the operators defined in Sec.~\ref{sec:Fermi_Liquid_from_commutations}, we define the operator $\hat F(x)$:
\begin{align}
F(\mathbf{y},\mathbf{p})=\epsilon^{ij}\frac{\partial}{\partial x^i}\delta(\mathbf{x}-\mathbf{y})p^j
, \quad 	\hat{F}(\mathbf x)=\int\! { d^2 \mathbf{y}\, \frac{d^2 \mathbf{p}}{4\pi^2}\,F(\mathbf{y},\mathbf{p})n_{\mathbf{p}}(\mathbf{y})},
\end{align}
and rewrite the density as:
\begin{equation} 		
\rho(\mathbf x)=\frac{B-b(\mathbf x)}{4\pi}-\frac{1}{B}\hat{F}(\mathbf{x}).
\end{equation}
We can now perform a Fourier transform and utilize commutation relations \eqref{FG-commut} to derive  the GMP algebra at leading order, Eq.~(\ref{eq:leading-GMP}), again taking into account the constraint~(\ref{eq:constr1}).

Note that for our derivation of the GMP algebra, the contribution of  the electric dipole moment of the Dirac composite fermion to the charge density is of crucial importance. 
We note that though, as far as we know, this is the first time that this algebra has been explicitly derived in any composite fermion model of FQHE, it can also be derived in the old dipolar model of Ref.~\cite{Stern:dipole} [see Eqs.(6) and (36) therein]. The GMP algebra also comes out naturally in a recently proposed ``bimetric'' theory of the nematic phase transition~\cite{Gromov:2017qeb}.


\section{Collective excitations}
\label{sec:dispersion}

In this section we look at the neutral excitations of the Dirac composite fermion near half filling. The methodology and the calculations of these sections were previously laid out in \cite{nguyenemIQH}. After a short review, we generalize the calculation to include a long range Coulomb interactions. We also compare our results to that of HLR \cite{Wang:HLR} and Jain's work~\cite{Scarola:2000}  qualitatively. We see that qualitatively our results fit quite well with experimental data \cite{Kukushkin:2009,Pinczuk:1993}.

We start with the recursion relation \eqref{eq:recur1} and  constraint equations \eqref{eq:constr1} and \eqref{eq:constr_k} and turn off external sources $A_0=0,\delta A_i=0$. We have:
%
\begin{subequations}\label{eq:recursion0}
\begin{multline}
\Big[\tilde{\omega}+\sign(b_0)n(1+F_n)\Big] u_n  = 
    \frac{B}{2|b_0|}\tilde{q}\Big[(1+F_{n+1})u_{n+1}+ (1+F_{n-1})u_{n-1}\Big] \\
    - \tilde{\omega}(\delta_{n,1}\delta a_z+\delta_{n,-1}\delta a_{\bar{z}}),
\end{multline}
\begin{equation}
u_{\pm1}=\pm\frac{\alpha}{4\frac{B}{|b_0|}(1+F_1)\omega_b \ell_B(1-\frac{|b_0|}{2B (1+F_1)\omega_b \ell_B}\alpha \tilde{q})}\tilde{q}(\delta a_z-\delta a_{\bar{z}}).
\end{equation}
\end{subequations}		
where using rotational symmetry have put $q_z=q_{\bar{z}}=\frac{q}{2}$ ($q_x=q,q_y=0$) and $\tilde{\omega}=\frac{\omega}{\omega_b}$ such that $u_n= u_n(\tilde q, \tilde w)$. Note also that since $u_0$ is nothing but the fluctuations of the composite fermion density, in the absence of deviations from the background magnetic field $\delta B=0$, equations \eqref{eq:unJ} and \eqref{eq:constr1} imply that $u_0 = 0$. It is also convenient to define rescaled momentum and interaction strength $z$ and $\lambda$: 
\begin{equation}
\frac{B}{|b_0|}\tilde{q}=z,\qquad
\lambda=\frac{\alpha}{4\frac{B^2}{b_0^2}\omega_b (1+F_1)\ell_B}.
\end{equation}  

In what follows we will assume that $\nu<\frac12$, that is we specialize to the case of $\nu = \frac{N}{2N+1}$.
The case of $\nu = \frac{N+1}{2N+1}$ follows in a similar fashion. We now replace $b_0=\frac{B}{2N+1}$ and $\sign(b_0)=1$. The above equations simplify as:
\begin{align}
\label{eq:recursion}
\Big[\tilde{\omega}+n(1+F_n)\Big]u_n=&\frac{z}{2}\Big[(1+F_{n+1})u_{n+1}+ (1+F_{n-1})u_{n-1}\Big]-\tilde{\omega}(\delta_{n,1}\delta a_z+\delta_{n,-1}\delta a_{\bar{z}}),\\
\label{eq:constraint1}
u_{\pm1}=&\pm\frac{\lambda z}{1-2\lambda z} (\delta a_z-\delta a_{\bar{z}}).
\end{align}

For simplicity, we now assume that the only nonzero Landau parameter is $F_1$, and $F_n=0$ for $n \neq \pm2$, although the technique we will described can be used there is any finite number of nonzero Landau parameters. 
Equation (\ref{eq:recursion}) is now a recursion relation, whose solution for $|n|>1$ with the requirement that $u_n\to0$ when $n\to\pm\infty$ is
\begin{subequations}\label{eq:Ansatz}
\begin{align}
u_n=&F(\tilde{\omega},z)J_{n+\tilde{\omega}}(z)   &n >& \,1,& \\
u_n=&G(\tilde{\omega},z)(-1)^n J_{-n-\tilde{\omega}}(z)   &n <& \!-\!1,&
\end{align}
\end{subequations}
where $F(\tilde{\omega},z)$ and $G(\tilde{\omega},z)$ are two unknown functions to be determined. Plugging this Ansatz into the recursion relation \eqref{eq:recursion}, we see that the equations for $n=\pm2,\pm1$ determine $u_{\pm1}$,  $\delta a_z$ and $\delta a_{\bar z}$:
\begin{subequations}
\begin{align}
u_1=&\frac{F(\tilde{\omega},z)}{1+F_1}J_{1+\tilde{\omega}}(z),\\
u_{-1}=&-\frac{G(\tilde{\omega},z)}{1+F_1}J_{1-\tilde{\omega}}(z),\\
\delta a_z=&\frac{F(\tilde{\omega},z)}{2\tilde{\omega}}\left[-\frac{2(1+F_1+\tilde{\omega})J_{1+\tilde{\omega}}(z)}{1+F_1}+zJ_{2+\tilde{\omega}}(z)\right],\\
\delta a_{\bar{z}}=& \frac{G(\tilde{\omega},z)}{2\tilde{\omega}}\left[-\frac{2(1+F_1-\tilde{\omega})J_{1-\tilde{\omega}}(z)}{1+F_1}+zJ_{2-\tilde{\omega}}(z)\right].
\end{align}
\end{subequations}
Finally, plugging these into the two constraint equations \eqref{eq:constraint1}, we arrive at a relationship between the undetermined functions $F(\tilde \omega,z)$ and $G(\tilde \omega,z)$: 
\begin{equation}
\label{eq:ratio}
\frac{F(\tilde{\omega},z)}{G(\tilde{\omega},z)}=\frac{ J_{1-\tilde{\omega}}(z)}{J_{1+\tilde{\omega}}(z)},
\end{equation}
as well as a final constraint, implicitly defining the dispersion relation $\tilde \omega(z)$ :
\begin{multline}
\label{eq:dispersion}
 J_{1-\tilde{\omega}}(z)=
\frac{\lambda z}{2(1-2\lambda z)\tilde{\omega}} \left\{ \frac{ J_{1-\tilde{\omega}}(z)}{J_{1+\tilde{\omega}}(z)}\left[-\frac{2(1+F_1+\tilde{\omega})J_{1+\tilde{\omega}}(z)}{1+F_1}+zJ_{2+\tilde{\omega}}(z)\right]\right.\\
 \left. -\left[-\frac{2(1+F_1-\tilde{\omega})J_{1-\tilde{\omega}}(z)}{1+F_1}+zJ_{2-\tilde{\omega}}(z)\right]\right\}.
\end{multline}
Note that if we turn off the long-range interaction ($\lambda=0$), $F_1$ drops out from the formulas and the dispersion relation simplifies to \cite{GNRS1}: 
\begin{align}
\label{eq:disnoCoulomb1}
&J_{1-\tilde{\omega}}(z)=0, \quad G(\tilde{\omega},z)=0,
\end{align}
which corresponds to $u_{n}=0$ for $n<0$, or 
\begin{align}
\label{eq:disnoCoulomb2}
&J_{1+\tilde{\omega}}(z)=0, \quad F(\tilde{\omega},z)=0,
\end{align}
which corresponds to $u_{n}=0$ for $n>0$.\\ 
\begin{figure}[h]
	\begin{subfigure}[h]{0.46\linewidth}
		\includegraphics[width=\linewidth]{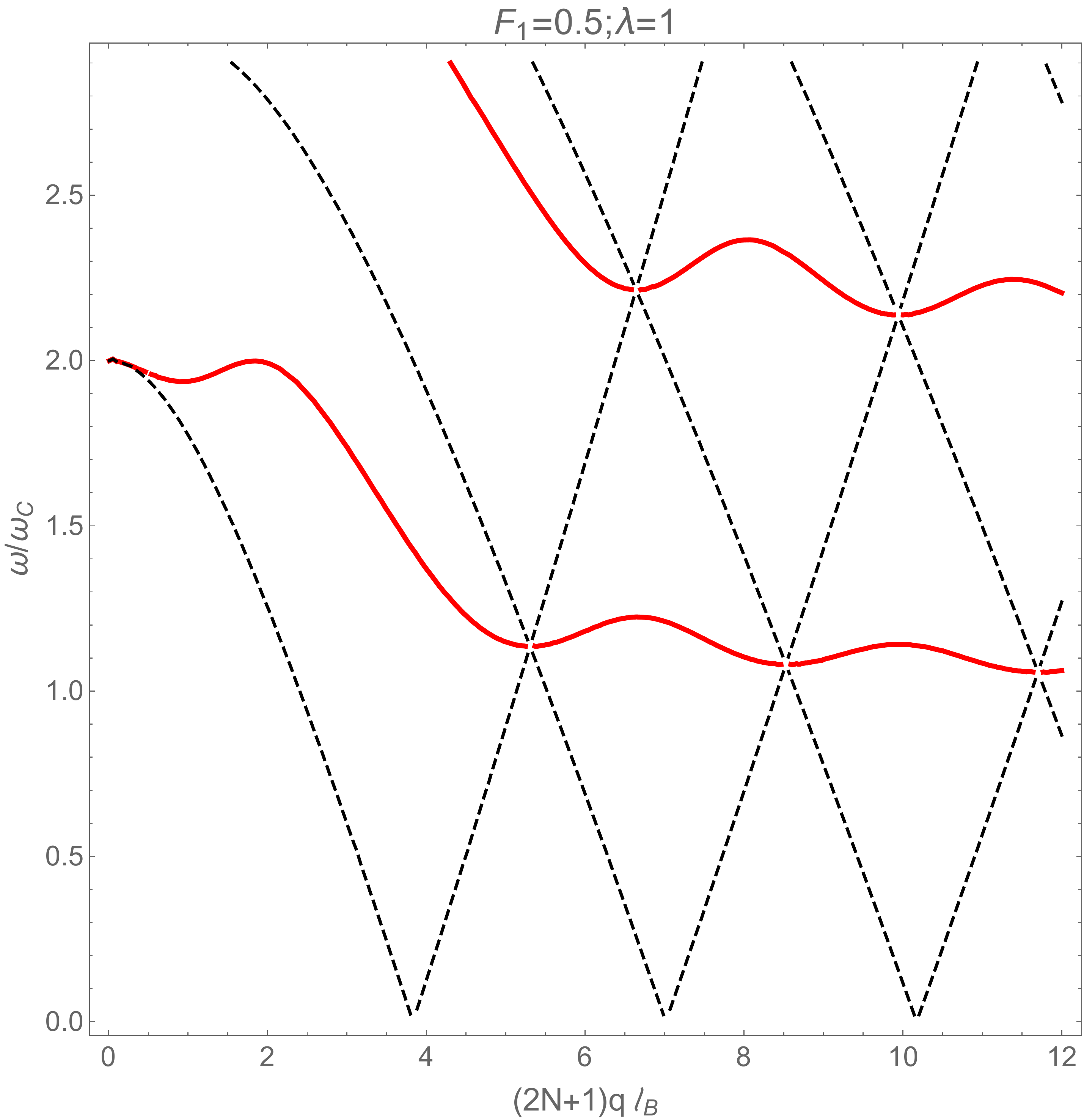}
		\caption{$F_1 = 0.5,~ F_{n>1}=0,~\lambda = 1$}
		\label{fig:41}
	\end{subfigure}
	\hfill
	\begin{subfigure}[h]{0.46\linewidth}
		\includegraphics[width=\linewidth]{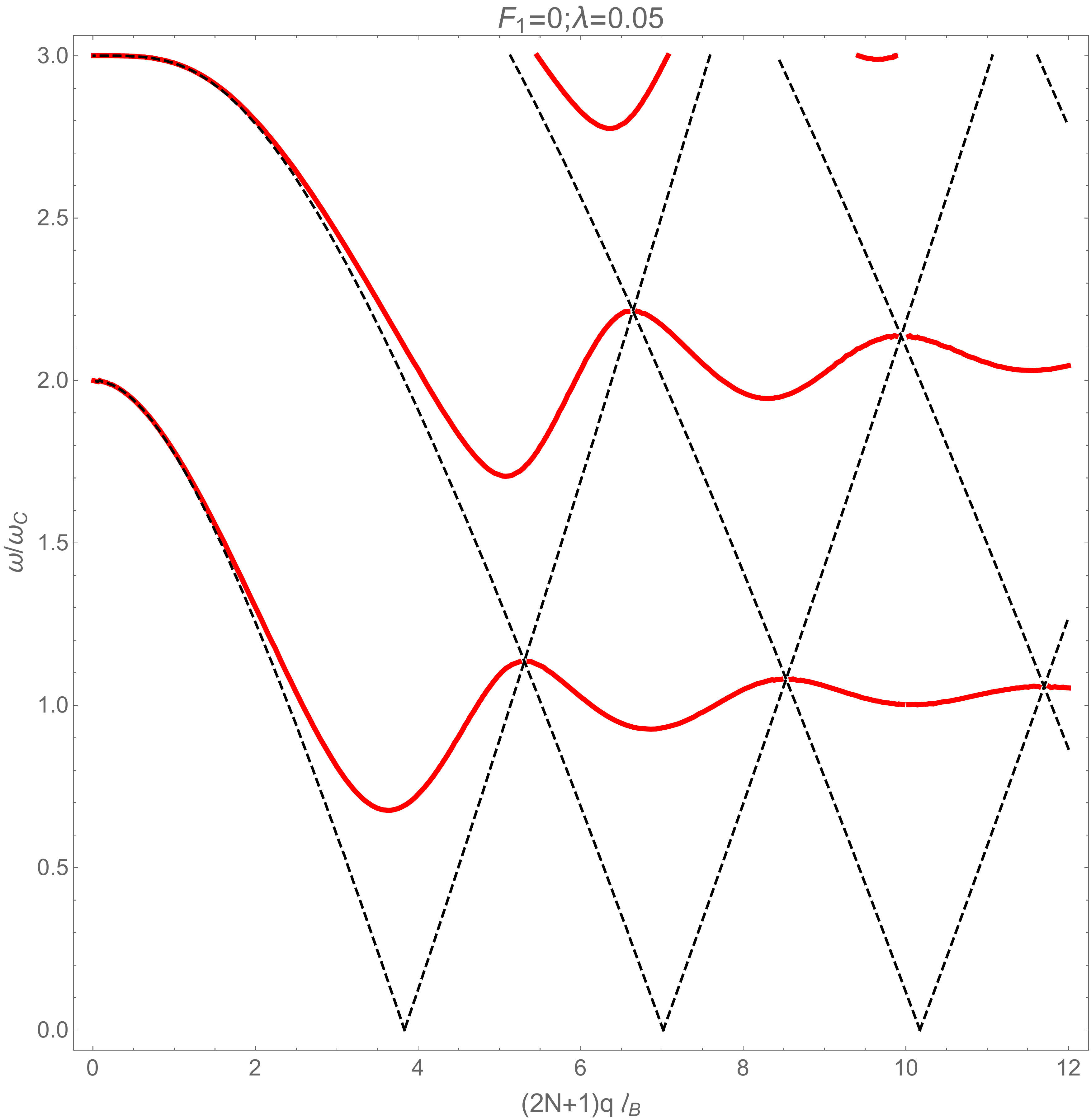}
		\caption{$F_{n}=0,~\lambda = 0.05$}
		\label{fig:00}
	\end{subfigure}
	\caption{Dispersion relation of neutral excitations of the Jain series $\nu = \frac12 \pm \frac1{2(2N+1)}$. The solid and dashed lines represent the dispersions with and without the Coulomb interactions. The $N$ dependence of these curves is only through the rescaling of the $x$ axis.}
	\label{fig:dispersion_examples}
\end{figure}
For given values of the Landau parameters $F_n$ and interaction strength $\lambda$ we can solve these equations numerically. The dispersion relation for two different cases can be seen in figure \ref{fig:dispersion_examples}. Note that the qualitative shape of these curves can depend strongly on the parameters. Here, we will comment on some of the more salient features of the dispersion relation.

\begin{enumerate}
	\item An important feature of the neutral excitations of the composite Dirac fermion theory is their particle-hole symmetry. We can show explicitly that the dispersion curves of the particle-hole duals $\nu_{\pm}=\frac{1}{2}\pm\frac{1}{2(2N+1)}$  are exactly equal, since the solution of \eqref{eq:recursion0} with $\sign(b_0)=-1$ can be obtained from the solution with $\sign(b_0)=1$ by the map 
	\begin{align}
u_n\leftrightarrow u_{-n}, \qquad \delta a_z \leftrightarrow \delta a_{\bar z}\,.
	\end{align}	
One may expect that the HLR theory, being not manifestly particle-hole symmetric, would give rise to different spectra of neutral excitations for the $\nu_-$ and $\nu_+$ states. However, this is not true, for a reason that we do not completely understand. In the Appendix \ref{sec:HLRDis}, we demonstrate that the dispersion relation in HLR theory \cite{Wang:HLR} is also PH symmetric up to next-to-leading order in $1/N$ expansion.  Furthermore, the HLR dispersion coincides with the dispersion derived in this section when the Coulomb interaction and Landau parameters are both turned off ($\lambda=0$ and $F_n=0$).  This seems to support the assertion made in Ref.~\cite{Wang:HLR} that the HLR theory has an emergent particle-hole symmetry in the IR. However, as we will see in Sec.~\ref{sec:Haldane}, the correlation functions computed from the HLR theory violate particle-hole symmetry.
	
	\item The energy gap for exciting any of the modes is independent of the interaction strength $\lambda$. This can be seen directly from the recursions relation and constraint equations \eqref{eq:recursion} and \eqref{eq:constraint1}, where the $\lambda$ dependence completely drops out at zero momentum $z=0$. Explicitly, the energy gap of the $n$th mode $\Delta_n$ is given by:
	\begin{equation}\label{eq:Delta_n}
	\Delta_n = (1+F_n)\omega_b\,.
	\end{equation} 
	
	\item  In the large-$N$ limit, the slope of the dispersion curve of the lowest mode at $q=0$ is generically negative. Solving Eqs.~\eqref{eq:recursion} and \eqref{eq:constraint1} perturbatively in the rescaled momentum parameter $z$, we see that the lowest mode has a dispersion which goes as:
	\begin{equation}
	\label{eq:omega2}
	\frac{\omega_2(q)}{\Delta_2} = 1-\frac{(2N+1)^2}{24(1-\Delta_2/\Delta_3)}q^2\ell_B^2+\cdots 
	\end{equation}
	where in deriving the above, we have assumed that the mode $n=2$ has the smallest gap. Note that this result is independent of the $\lambda$, the strength of the Coulomb interaction. 

\item An interesting feature of the dispersion relation of the neutral excitation is the dependence of the relative height of the first two minima on the strength of the Coulomb interaction. For example for $F_1 = 0.5$ and $\lambda = 1$ (figure \ref{fig:41}) the first minima of each mode are higher than the second minima. The situation is reversed when $F_1 = 0$ and $\lambda = 0.05$ (figure \ref{fig:00}). Generally when $\lambda$ is small, we see the first minima are lower than the second and the height of the minima flip when $\lambda$ becomes of order one. There is also an intermediate range of values for $\lambda$ where some of the minima disappear. At large $N$, $\lambda\sim N^{-1}$, so the first minimum is lower than the second one.
	
The most precise  measurement of the dispersion of the neutral excitations was done by Kukushkin \emph{et al.}\ in 2009 \cite{Kukushkin:2009}. Looking at Fig.~4 in that paper~\footnote{Note that the different curves in Fig.~4 of Ref.~\cite{Kukushkin:2009} are the dispersions of the lowest neutral excitation for three different filling fractions, whereas the curves in Fig.~\ref{fig:dispersion_examples} are the dispersions of the first three lowest excitations at the same filling fraction.}, it would appear that the second minimum of the dispersion curve is lower than the first~\footnote{It is worth pointing out that the measurements of Ref.~\cite{Kukushkin:2009} are not sufficiently precise to definitely determine which minimum is lower.}. Our model can give this kind of behavior only for large $\lambda$ and moderate $F_1$.  
	
\end{enumerate}

\section{Electromagnetic response}
\label{sec:response}

In this we look at the the electromagnetic response functions of the composite Dirac fermions in Jain's states $\nu_\pm=\frac12\pm\frac{1}{2(2N+1)}$,  using the semiclassical approach laid out in Sec.~\ref{sec:semiclassical}. We will then explicitly check the particle-hole symmetry of the results and compare to the results from HLR and Jain's theories. 

We again start with the recursion relation \eqref{eq:recur1} and the constraint equations \eqref{eq:constr_k}. The computation follows just as in Sec.~\ref{sec:dispersion}, however in this section, we turn on the external field $\delta A_\mu$ in order to compute the response functions. Explicitly, if we assume $F_2$ to be the only nonzero Landau parameter, we again assume the Ansatz \eqref{eq:Ansatz} for $u_{|n|>2}$ and solve for the remaining fields $u_{\pm 1}$, $a_z$, $a_{\bar{z}}$ as well as the unknown functions $F(\tilde\omega,z)$ and $G(\tilde\omega,z)$ by substituting the Ansatz into the recursion and constraint equations. The case with more Landau parameters turned on is a slight generalization of the above.

Finally, we read off the electromagnetic response of the system from equations \eqref{eq:J} which give the electromagnetic density and currents in terms of external field $\delta A_\mu$. The polarization tensor $\Pi^{\mu\nu}$ which encodes the linear response of the system to the EM fields is defined as:
\begin{equation}
\delta J^\mu(\omega,q)=\Pi^{\mu\nu}(\omega,q)\delta A_\nu(\omega,q),
\end{equation}
where $\delta J^\mu=J^\mu-\bar{J}^\mu$ is the deviation of the currents from their values at equilibrium given by $\bar{\rho}=\frac{\nu}{2\pi}B$ and $\bar{J}^i=0$.

The explicit calculation can be derived from a straightforward modification of the presentation in Sec.~\ref{sec:dispersion}. We therefore skip it and quote the results. Note that using this method, it is possible to derive the response functions in closed analytic form if there is only a finite number of nonzero Landau parameters.
Here, we will only report the first few terms in a gradient expansion in small momenta $q$.

\subsection{Susceptibility}

We first look at the susceptibility of the system, which is given by:
\begin{equation}
\chi(\omega,q)=\Pi^{00}(\omega,q)=\frac{\delta}{\delta A_0} \rho(\omega,q)\Big|_{\delta A_\mu = 0},
\end{equation}
which encodes the response of the electric charge density to variations of the scalar potential $\delta A_0$. 
We can evaluate this susceptibility in power expansion over $q$.
We see that to order $\cO(q^6)$ the susceptibility for the $\nu_+$ and $\nu_-$ states depends only on the two Landau's parameters $F_2$ and $F_3$ and
is given by
\begin{multline}\label{eq:chipm}
\chi_+(\omega,q) = \chi_-(\omega,q)=-\frac{q^4\ell^2(2N+1)\omega_2(q)}{32\pi(\omega^2-\omega^2_2(q))} \\ +\frac{(2N+1)^3q^6\ell^4}{768\pi}\left\{\left[1-\frac{\Delta_2^2}{(\Delta_3{-}\Delta_2)^2}\right]  \frac{\Delta_2}{\omega^2-\Delta_2^2}
   +\frac{\Delta_2^2}{(\Delta_3{-}\Delta_2)^2}\frac{\Delta_3}{\omega^2-\Delta_3^2}\right\},
\end{multline}
where $\omega_2(q)$ is given in Eq.~\eqref{eq:omega2} and $\Delta_{2,3}$ by Eq.~(\ref{eq:Delta_n}).

\subsection{Projected static structure factor and the Haldane bound}
\label{sec:Haldane}

A quantity of interest closely related to the susceptibility is the projected statistic structure factor $\bar s(q)$. It can be evaluated using the equation \cite{Girvin:1986zz,NS}:
\begin{equation}
\bar{s}(q)=-\frac{i}{\rho}\int \frac{d\omega}{2\pi} \chi(\omega+i\,\epsilon\, \sign(\omega),q)
=-\frac{1}{\rho}\sum_{\omega_i>0}\text{Res}(\chi(\omega,q),\omega=\omega_i),
\end{equation}
where $\epsilon$ is an infinitesimal positive number. We find
\begin{equation}
\label{eq:SSFh}
\nu_\pm\bar{s}_\pm(q) = \frac{ (2 N+1) q^4\ell_B^4}{32}+\frac{ (2 N+1)^3 q^6\ell_B^6}{768 }
+\cO(\alpha q^7).
\end{equation}
Let us $s_n$ through the Taylor expansion $\bar s = \sum s_n (q \ell_B)^n$.  We see that $s_4$ and $s_6$ are independent of the interaction parameters $F_n$ and $\lambda$.  The dependence on Landau parameters first appears in the $q^7$ term.
In a previous work \cite{Nguyen:2016ph}, we determined that under certain assumptions, $s_4$ and $s_6$ can be related to various topological properties of the system~\footnote{The assumptions of Ref.~\cite{Nguyen:2016ph}, can be summarized as the requirement of Galilean invariance and that the ground state satisfying certain chirality properties.}. Here, we explicitly verify these claims. In particular, we find that $s_4$ is determined by the Wen-Zee shift $\cS$:
\begin{equation}
  s_4^\pm = \mp \frac{\cS_\pm-1}8,
\end{equation}
where $\cS_+ = -N + 1 $ and $\cS_- = N + 2$ are the shifts of the $\nu_+$ and $\nu_-$ states, respectively. This suggests that the states $\nu_-$ and $\nu_+$ are chiral and anti-chiral states in the language of Ref.~\cite{Nguyen:2016ph}, respectively: in these states, the Haldane bound~\cite{Haldane:bound1} $s_4\geq \frac18|\cS-1|$ is saturated. As explained in Ref.~\cite{Nguyen:2016ph}, for the Haldane bound to be saturated, it is sufficient for excited states at zero momentum to carry angular momentum of the same sign. For Jain's states, this condition is satisfied thanks to the fact that among $u_n$, the creation operators have the same sign of $n$ while the annihilation operators have the opposite sign.


These results can be compared with those obtained within the HLR theory.
In the MRPA scheme of Ref.~\cite{Simon:1993} (see also the Appendix of Ref.~\cite{SonLevin:PH}) one can calculate the electromagnetic response from the polarization tensor of free fermion in a magnetic field.  One obtains, after straightforward algebra
\begin{subequations}
\begin{align}
  \chi_- (\omega,q) &= -\frac{q^4\ell_B^2}{4\pi} \frac{\omega_b}{\omega^2-4\omega_b^2}\, \frac{N^2}{2N+1}\,,\\
  \chi_+(\omega,q) &= - \frac{q^4\ell_B^2}{4\pi} \frac{\omega_b}{\omega^2-4\omega_b^2}\, \frac{(N+1)^2}{2N+1}\,.
\end{align}
\end{subequations}
The resulting expressions differ from those obtained in the Dirac composite fermion [Eq.~(\ref{eq:chipm})] theory by $\cO(1/N)$.  More crucially, $\chi_-$ and $\chi_+$ do not coincide, as required by particle-hole symmetry. Computing the static structure factors, we find
\begin{subequations}\label{eq:s4-HLR}
\begin{align}
  s_4^- & = \frac N8\,,\\
  s_4^+ & = \frac{N+1}8\,.
\end{align} 
\end{subequations}
Thus, in the HLR theory, the Haldane bound is only satisfied for the $\nu_+$ states and is violated for $\nu_-$ states.

 What is the reason for the violation of the Haldane bound in HLR theory?
The lower bound on $s_4$ is a consequence of Galilean invariance and the LLL projection  \cite{Haldane:bound1,Haldane:bound2,Golkar2016}.
Since the MRPA scheme preserves Galilean invariance, the natural conclusion is that the HLR theory is not consistent with the LLL projection.
To shed light on the violation of the Haldane bound in the HLR theory, we recall a derivation of this bound in Ref.~\cite{Golkar2016}, where two spectral sum rules were shown to hold:
\begin{align}
  \int\limits_0^\infty\! \frac{d\omega}{\omega^2}\, [\rho_T(\omega)-\bar\rho_T(\omega)] &= \frac{\eta_H(0)-\eta_H(\infty)}{2\rho}\,, \\
    \int\limits_0^\infty\! \frac{d\omega}{\omega^2}\, [\rho_T(\omega)+\bar\rho_T(\omega)] &= s_4\,.
\end{align}
Here $\eta_H(0)$ is the Hall viscosity at zero frequency, and $\eta_H(\infty)$ is the Hall viscosity at frequency much larger than the Coulomb energy scale, but much smaller than the cyclotron energy (which is infinite in the LLL limit), $\rho_T(\omega)$ and $\bar\rho_T(\omega)$ are the spectral densities of the holomorphic and antiholomorphic components of the stress tensor. From the positivity of the spectral densities one obtains a lower bound
\begin{equation}
  s_4 \ge \frac1{2\rho} |\eta_H(0)-\eta_H(\infty)|,
\end{equation}
which is saturated when one of the spectral functions vanishes, as for chiral states. When one takes $\eta_H(\infty)=\frac12\rho s(\infty)$ with $s(\infty)=\frac12$, Haldane's bound follows. Here $s(\infty)$ may be interpreted as the spin per particle at energy much larger than the interaction energy scale.
As argued in Ref.~\cite{SonLevin:PH}, the HLR theory assigns an incorrect (particle-hole asymmetric)  value for $s(\infty)$, equal to half the number of flux quanta attached to each composite fermion: $s(\infty)=1$.  As the result, for chiral states the HLR theory predicts
\begin{equation}
  s_4(\textrm{HLR}) = \frac18 |\mathcal S-2|,
\end{equation}
which reproduces Eqs.~(\ref{eq:s4-HLR}). Thus, the violation of the Haldane bound by the HLR theory can be traced back to the wrong orbital spin it assigns to the composite fermion.

We also find that the quantity $s_6$ is determined completely by the filling fraction $\nu$, the shift $\cS$, the chiral central charge $c_-$ \cite{KITAEV}  and  the orbital spin variance $\text{var}(s)$ \cite{Gromov:vars,Bradlyn:vars} in the manner described previously in Ref.~\cite{Nguyen:2016ph}. 

\subsection{DC Hall conductivity}

The Hall conductivity is defined as:
\begin{equation}
\sigma^H(\omega,q)=\frac{\Pi^{12}(\omega,q)}{i\omega}=\frac{\delta}{\delta A_1} J^2(\omega,q)\Big|_{\delta A_\mu = 0},
\end{equation}
In Coulomb gauge, where $E_i(\omega,q)=i\omega \delta A_i(\omega,q)$, this equation takes on the familiar form   
\begin{equation}
J^1(\omega,q)=\sigma^H(\omega,q) E_2(\omega,q),
\end{equation}
 which is the current density in term of the perpendicular applied electric field. 
The DC Hall conductivity is  the limit of the Hall conductivity as the frequency goes to zero:
\begin{equation}
\sigma^H(q)=\lim_{\omega \rightarrow 0}\sigma^H(\omega,q).
\end{equation}
In the case of conductivities we do not expect the result for $\nu_+$ and $\nu_-$ to be the same, even in the presence of particle-hole symmetry. We have:
\begin{equation}
\label{eq:Hall_Response}
\sigma^H_\pm(q)=\frac{N+\frac12 \pm \frac 12}{2\pi(2N+1)} \mp \frac{4N^2+2N+3\pm(2N+2)}{32\pi(2N+1)}q^2\ell_B^2+\cO(\alpha q^3).
\end{equation} 
The DC Hall conductivity satisfies the relation:
\begin{equation}
\label{eq:Shift1}
\sigma^H_\pm (q)\approx \frac{\nu_\pm}{2\pi}
\Bigl(1+\frac{ \cS_\pm-2}{4}q^2\ell_B^2\Bigr)+\cdots
\end{equation}
where $\approx$ means equal up to next-to-leading in $1/N$ expansion and $\cS$ is the Wen-Zee shift given by $\cS_+ = -N + 1 $ and $\cS_- = N + 2$ \cite{NS,Nguyen:2016ph,Son:dirac2016}.

Note that both the $q^0$ and $q^2\ell_B^2$ coefficients of DC Hall conductivity are purely  topological numbers, determined solely by the filling fraction $\nu$ and the shift $\cS$. These results were first derived~\footnote{Similar results were derived in  Refs.~\cite{Nguyen:2016ph,Paul:wavefn1} using a wave-function approach. } in Refs.~\cite{son2013newton,sonhoyos,Read:Hallk2} for generic fractional quantum Hall states using nothing but Galilean invariance~\footnote{In Ref.~\cite{son2013newton}, the Hall conductivity is computed for generic electron $g$-factor. Setting $g=2$ gives the result (\ref{eq:Shift1}). In Refs.~\cite{sonhoyos,Read:Hallk2} the authors presented the result of  Ref.~\cite{son2013newton} for $g=0$. }. This is a nontrivial consequence of the presence of Galilean invariance in the Dirac composite fermion model.

Note that the inequality of $\sigma^H_+$ and $\sigma^H_-$ is to be expected. Indeed, the naive expectation from particle-hole symmetry  would be that  the sum of the conductivities of the particle-hole conjugate states  $\nu_+$ and $\nu_-$ be equal to the conductivity of the full Landau level. In momentum space this relation takes the form
 \begin{equation}
\sigma^H_+(\omega,\mathbf{q})+\sigma^H_-(\omega,\mathbf{q}) = \frac{1}{\pi}\frac{1-e^{-q^2\ell_B^2/2}}{q^2 \ell_B^2}.
\end{equation}
However, it was shown that in the presence of interactions, the above equality is modified~\cite{SonLevin:PH} and we have
 \begin{equation}
 \label{eq:PH_Hall_relation}
 \sigma^H_+(\omega,\mathbf{q})+\sigma^H_-(\omega,\mathbf{q})+\frac{1}{2\pi}\tilde{V}(\mathbf{q})\chi(\omega,\mathbf{q})
  = \frac{1}{\pi}\frac{1-e^{-q^2\ell_B^2/2}}{q^2 \ell_B^2},
 \end{equation}
 where $\chi(\omega,q)$ is the susceptibility as defined above and $\tilde V(q)$ is fully determined by the electron-electron interaction. In the case of the Coulomb interaction $\tilde V(q)$ takes is given by~\cite{SonLevin:PH}
 \begin{equation}
 \tilde{V}(\mathbf{q})=4\pi\alpha\biggl(\frac{1-e^{-q^2\ell_B^2/2}}{q^3\ell_B^2}-\frac{1}{q^2\ell_B}\sqrt{\frac{\pi}{2}}\Big[1-e^{-q^2\ell_B^2/4}I_0\left(\frac{q^2\ell_B^2}{4}\right)\Big]\biggr),
 \end{equation}
 where $I_0$ is the modified Bessel function of the first kind. 
 
 As we have started from the effective Lagrangian constructed by taking into account the prescription of Ref.~\cite{SonLevin:PH}, which explicitly sought to satisfy constraints of the type (\ref{eq:PH_Hall_relation}), our result should satisfy this constraint.  
 We have performed a test for \eqref{eq:PH_Hall_relation}
 for a particular case when the only nonzero Landau parameter is $F_2$.
 In this case $\sigma_\pm^H$ can be computed exactly in terms of Bessel functions, thought the expressions are rather cumbersome.
We have explicitly checked that relationship~(\ref{eq:PH_Hall_relation}) is indeed satisfied up to and including the next-to-leading order in $1/N$. This, along with the equality of the susceptibilities \eqref{eq:chipm}, provide us with two explicit verifications of particle-hole symmetry in the Fermi liquid approach, even in the presence of Landau parameters $F_n$. We note that HLR and modified HLR theory pass neither of these two tests of particle-hole symmetry. 


\section{Conclusion}
\label{sec:conclusion}

In this paper, we have studied the electromagnetic response of quantum Hall systems in the Dirac composite  fermion theory. We performed a semiclassical calculation to obtain closed form results in the long wavelength limit. The results demonstrate explicitly that the PH symmetry is present in both response functions and dispersion relations of excited states. Our calculation of the dispersion relation shows qualitative agreement with experimental results  in both the positions of the minima and the trend of the dispersion relation curve. 

 We explicitly confirmed the PH duality relations of response functions of Ref.~\cite{SonLevin:PH} in the presence of Coulomb interactions. The particle hole symmetry of our results is indeed expected as we start with a  manifest PH symmetric theory. From our analytical results, we reproduce the topological quantum numbers of Jain sates, matching previous work \cite{Nguyen:2016ph}. 
 
We have compared our results with the outcome of HLR theory \cite{Wang:HLR}, showing that HLR theory does not satisfy PH symmetry in electromagnetic response functions, resulting in a sharp distinctions between Dirac composite fermions and HLR. Nevertheless, the dispersion relation of neutral excitation, computed from the HLR theory, is particle-hole symmetric to leading and next-to-leading order in $1/N$. Is it possible that the HLR theory can be modified, e.g., by adding extra terms to the Lagrangian which contains the external gauge field, to restore particle-hole symmetry? It would be extremely interesting possibility, though currently we do not have any concrete proposal. It seems that the incorrect value of the high-frequency Hall viscosity in the HLR theory, identified in Ref.~\cite{SonLevin:PH}, is the first issue one needs to resolve.
 
We also derived the GMP algebra from our effective field theory picture and demonstrated that the crucial role of the electric dipole moment of the composite fermion. The presence of the GMP algebra is a signature of the lowest Landau level projection in a theory. In addition to PH symmetry, we consider the reproduction of both topological quantum numbers and GMP algebra as a nontrivial independent check for the validity of Dirac composite fermion as an effective field theory of the FQH. 


Finally, we note that the bosonized approach used in this paper is sufficiently flexible to accommodate any value of Landau's parameters.  In particular, it can be used to investigate the nematic phase transition, where $F_2\to-1$~\cite{BimetricCFL}.

\acknowledgments

It is a pleasure to thank Bert Halperin, Steve Simon, and Chong Wang for discussions.  This work is supported, in part, by DOE grant DE-FG02-13ER41958 and a Simons Investigator grant from the Simons Foundation.  Additional support was provided by the Chicago MRSEC, which is funded by NSF through grant DMR-1420709.
S.G. is supported by the James Arthur Postdoctoral Fellowship.


	\appendix

\section{Dispersion relation of HLR theory}
\label{sec:HLRDis}
In this appendix, we show that in the absence of Coulomb interaction, the dispersion relation of HLR theory \cite{Wang:HLR} coincides with the one in Dirac composite Fermi-liquid theory up to next-to-leading order in $1/N$ expansion. Starting from Eq.~(79) of Ref.~\cite{Wang:HLR}, we can extract the relationship between the frequency and momentum for $\nu_-$ as
\begin{multline}
\label{eq:HLRDis1}
\frac{8 \pi N\tilde\omega}{\sin\pi \tilde\omega} \Big[2 N J_{1-\tilde{\omega}}(X_-) J_{\tilde{\omega}+1}(X_-)+J_{-\tilde{\omega}}(X_-) (X_- J_{\tilde{\omega}+1}(X_-)-\tilde{\omega} J_{\tilde{\omega}}(X_-))\Big]\\+8 N \tilde{\omega}+X_-^2=0,
\end{multline}
where $X_-=z \sqrt{\frac{2N}{2N+1}}$, and the relation for $\nu_+$ as 
\begin{multline}
\label{eq:HLRDis2}
\frac{8 \pi(N+1)\tilde\omega}{\sin\pi\tilde\omega} \Big[2 (N+1) J_{1-\tilde{\omega}}(X_+) J_{\tilde{\omega}+1}(X_+)+J_{-\tilde{\omega}}(X_+) (\tilde{\omega} J_{\tilde{\omega}}(X_+)-X_+ J_{\tilde{\omega}+1}(X_+))\Big]\\-8 (N+1) \tilde{\omega}+X_+^2=0,
\end{multline}   
where $X_+=z \sqrt{\frac{2(N+1)}{2N+1}}$. Note the similarity between Eqs.~\eqref{eq:HLRDis1} and \eqref{eq:HLRDis2} and Eq.~\eqref{eq:dispersion} in Sec.~\ref{sec:dispersion}. Expanding Eq.~\eqref{eq:HLRDis1} in $1/N$ and keeping the leading and next-to-leading terms, we arrive at 
\begin{equation}
\label{eq:HLRDis3}
  \frac{8\pi N (2 N+1) \tilde\omega}{\sin\pi\tilde\omega} J_{1-\tilde{\omega}}(z) J_{\tilde{\omega}+1}(z)=0.
\end{equation}	
Similarly for Eq.~\eqref{eq:HLRDis2} we obtain 
\begin{equation}
\label{eq:HLRDis4}
\frac{8\pi N(2 N+3)\tilde\omega}{\sin\pi\tilde\omega} J_{1-\tilde{\omega}}(z) J_{\tilde{\omega}+1}(z)=0.
\end{equation}
Equations \eqref{eq:HLRDis3} and \eqref{eq:HLRDis4} give rise to exactly the same spectrum, determined the solutions to the equations $ J_{\tilde{\omega}+1}(z)=0$ and $ J_{1-\tilde{\omega}}(z)=0$. 
As equations for $\tilde\omega$ the solutions to these equations come in pairs of opposite signs. These equations coincide with \eqref{eq:disnoCoulomb1} and \eqref{eq:disnoCoulomb2} 
in the absence of both the Coulomb interaction ($\lambda=0$) and the Landau parameters ($F_n=0$).
We therefore see that the dispersion relation of HLR theory is PH symmetric and equal to the dispersion derived from composite Dirac fermions found in Sec.~\ref{sec:dispersion}.

	\bibliography{CFFL2}
	
\end{document}